\documentclass[acmsmall,screen]{acmart}

\usepackage{amsmath}
\usepackage{multirow}
\usepackage{array}
\usepackage{subfigure}
\usepackage{makecell}  % Include this in your preamble
\usepackage{algorithm}
\usepackage{algpseudocode, caption}
\usepackage{pseudocode}
\usepackage{graphicx}

\usepackage{subcaption}
\usepackage{mdframed}
\usepackage{caption}
\usepackage{algorithm}
\usepackage{algpseudocode}
\usepackage{listings}
\usepackage{enumitem}
\usepackage{soul}
\definecolor{darkgreen}{rgb}{0.0, 0.6, 0.0}
\AtBeginDocument{%
  }

\begin{document}

\title[LLM-Aided Customizable Profiling of Code Data Based On Programming Language Concepts]{LLM-Aided Customizable Profiling of Code Data Based On Programming Language Concepts}

\author{Pankaj Thorat}
\affiliation{%
  \institution{IBM Research}
  \city{Bangalore}
  \country{India}}
\email{pankaj.thorat@ibm.com}

\author{Adnan Qidwai}
\affiliation{%
  \institution{IIIT Hyderabad}
  \city{Hyderabad}
  \country{India}}
\email{adnan.qidwai@students.iiit.ac.in}

\author{Adrija Dhar}
\affiliation{%
  \institution{NIT Durgapur}
  \city{Durgapur}
  \country{India}}
\email{ad.21u10475@btech.nitdgp.ac.in}

\author{Aishwariya Chakraborty}
\affiliation{%
  \institution{IBM Research}
  \city{Bangalore}
  \country{India}}
\email{aishwariya.chakraborty1@ibm.com}

\author{Anand Eswaran}
\affiliation{%
  \institution{IBM Research}
  \city{Bangalore}
  \country{India}}
\email{anand.eswaran@ibm.com}

\author{Hima Patel}
\affiliation{%
  \institution{IBM Research}
  \city{Bangalore}
  \country{India}}
\email{himapatel@in.ibm.com}

\author{Praveen Jayachandran}
\affiliation{%
  \institution{IBM Research}
  \city{Bangalore}
  \country{India}}
\email{praveen.j@in.ibm.com}

\renewcommand{\shortauthors}{Thorat et al.}

\begin{abstract}
Data profiling, in the context of machine learning, is the process of examining and analyzing data to create useful statistics. These statistics are used both as an aid for better comprehension of the properties of data as well as for a variety of downstream data processing tasks such as data valuation (assessing the value of data relative to the business objectives at hand) and data curation (filtering and prioritizing training data based on derived thresholds). In the Large Language Model (LLM) setting, training data is typically unstructured in nature comprising natural language text, images, and code. In this work, we specifically focus on code-LLMs, where the quality of code training data substantially affects the model accuracy of LLM-based coding tasks such as code generation and summarization. Therefore, having the capabilities to characterize code data in terms of programming language concepts aids in both deriving insights related to code training/evaluation data and in the downstream curation of code training data. 
In this work, we address the problem of profiling multi-lingual code datasets by extracting an extensible user-defined set of syntactic and semantic concepts over arbitrary programming languages. The key novelty in our approach is the decomposition of the code data profiling problem into two phases — (1) a frugal offline phase, in which LLMs are used to derive and learn language-specific rules for extracting syntactic and semantic concepts from code snippets across arbitrary unknown programming languages, and (2) a deterministic online phase where simple language-specific rules are applied to analyze and categorize each code data sample to extract above-mentioned concepts. The set of concepts defined in our framework is extensible and customizable, thereby making them amenable to use-case specific specialization. Our hybrid approach is practical and can support rules for a large (21), diverse set of programming languages and a rich customizable range of semantic constructs. Our LLM-aided methodology exhibits a mean accuracy of 90.33\% for syntactic concept extraction rules across syntactic constructs and languages, and exhibits a mean semantic classification accuracy of 80\% and 77\% over languages and semantic concepts respectively.

\end{abstract}





\maketitle

\section{Introduction}
In recent years, advancements in Large Language Model (LLM) technology have sparked the creation of numerous innovative applications, enabling the development of new businesses and the enhancement of existing workflows. While some of these applications are built using LLMs out of the box, a large number of them necessitate the customisation of LLMs to a given use case. Popular ways of customisation include fine-tuning LLMs, instruct tuning LLMs, building Retrieval Augmented Generation (RAG) applications etc.  
%Let us take fine tuning as an example method of customisation for our discussion. 
A user's goal for fine-tuning an LLM could be to improve a chosen base LLM's performance for a specific task or language, by training it using additional proprietary data samples that the base model may have not seen. To make the discussion concrete, we will discuss the scenario where a user wants to improve a code LLM model's performance on a language of choice, like Verilog and has access to some Verilog data sets. At this point, important considerations related to the data set come up, such as, \textit{What languages are in these data sets and how much of it is indeed Verilog? Do they contain examples of all the relevant libraries from the Verilog language? Is the quality of the Verilog code samples high? Are samples in my Verilog data set similar to the code samples in the eval set? }
%Does one data set contain samples of libraries that aren't found in other data sets that make it unique and valuable? What functional use cases are supported by this data?

\begin{figure}[h]
    \centering
    \includegraphics[scale=0.3]{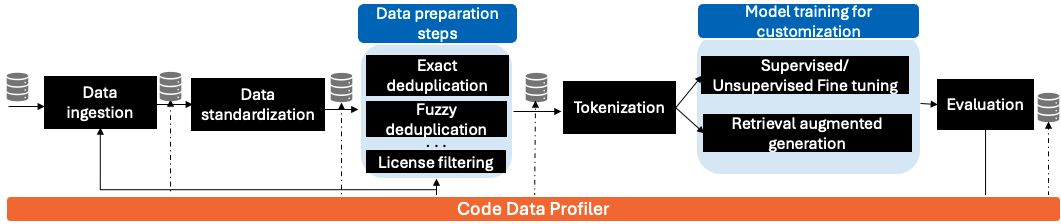}
    %\vspace{-10pt}
    \caption{Proposed Code Profiler Step that is Useful Across Data and Model Lifecycle.}
    \label{fig:intro}
\end{figure}

 %Without a detailed understanding of the data, users are essentially navigating the contents of these data sets blindly. 
 Without a detailed understanding of the data, users are ill-equipped to answer such questions, which impacts their model customization objectives. Coarse metrics like volume of data, while easy to measure, are insufficient for answering the above questions. For instance, a user may have 5000 new repositories containing Verilog code, but 80 percent of the code may only cover 5 percent of the total libraries! Once the right data is identified, the data usually undergoes some data preparation steps like exact deduplication, code quality filtering etc ~\cite{mishra2024granite}. Each of these steps analyses the quality of data and then cleans it by either changing the content or dropping code files. Thus, as the contents of the data set change across these steps, all the above questions remain relevant at each step of the data/model cycle. 
 %As the user goes further and trains the models, there is also a need to understand the composition of the evaluation set and be able to pinpoint libraries where the fine-tuned code model makes the most errors. 
 Answering such questions requires the foundational ability to analyze and expose code data statistics based on flexible user-defined criteria e.g. Verilog files in the data set that are both high-quality and well-documented. To address use cases similar those mentioned above, we develop and present a code data profiling tool that can help across the end-to-end data and model life-cycle as shown in figure~\ref{fig:intro}. 

Data profiling is a well-known module as part of data and model lifecycle and has been applied extensively for tabular datasets \cite{abedjan2015profiling}. In our work, we discuss code data profiling, where we also introduce new dimensions that are meaningful for data profiling for code datasets. We classify these dimensions into two types: \textit{syntactic concepts} and \textit{semantic concepts}. Syntactic concepts are based on definitions from programming languages like packages, modules, etc., and semantic concepts guide the user on functional use cases that can be enabled from given code data, e.g., code can support database interactivity. The challenge with code datasets as opposed to tabular datasets is the vast variety of programming languages that exist. According to Github \cite{octoverse:top_programming_languages} there are more than 500 active languages used in github projects today. For a code profiler to be useful, it should be able to natively support a large number of programming languages without any human intervention. 

Existing tools, such as Tree-sitter \cite{treesitter}, have already established a solid foundation for building abstract syntax trees (ASTs) across various programming languages. Nevertheless, the limitation arises when a user must develop a binding for each language, which means that the tool is not applicable across languages without human intervention. Furthermore, the tool can only support AST-based syntactic concepts, restricting its capabilities for partitioning and analyzing code data using use-case specific semantic profiling. % and more complex syntactic concepts such as average-function-length, which are used for customizable concept-based \textit{"slicing"} (partitioning) of code data and exposure of slice-level data statistics to the user. 
In this paper, we approach this problem of building a generic code data profiling tool that can work across a large number of programming languages using the power of LLMs and can profile against both syntactic and semantic concepts. Multi-lingual code metadata in our tool is represented in a uniform language-agnostic tabular scheme that we call the \textit{Uniform Base Syntactic Representation} (UBSR) enabling flexible SQL-like querying and concept composability. We are also sensitive to the costs incurred by using LLMs, so we propose a novel approach to building this tool by breaking it into an \textit{online phase} and an \textit{offline} phase. The offline phase is a one-time phase where we use LLMs to generate rules for the profiling concepts of interest, and the online phase can use these rules to profile any new code snippets. While it can be argued that such rules can be handwritten, this requires a developer to be adept across a large number of programming languages, which is cumbersome and time-consuming. The main contributions of this paper are: 

\begin{itemize}[leftmargin=*]
\item A new code profiler tool that can be used to profile code data and derive statistics, which is generalizable to a large number of programming languages \textit{without any human intervention}.
\item Propose a new unified structured syntactic representation of code that enables the code profiling tool to be multilingual. 
\item A cost-sensitive design where we use LLMs to generalise across programming languages without incurring the cost for every input sample at run time.
\item A flexible architecture that allows a user to adapt the semantic concepts as per the use case. 
\end{itemize}

 The rest of this paper is organized as follows: Section \ref{refwork} discusses relevant related work. Section \ref{sol-outline} outlines our solution, Section \ref{sysarchsec} covers our design in depth while Section \ref{impl} discusses implementation details. Section \ref{expsec} examines our experimental results while Section \ref{conclusion} summarizes our contributions, outlining future work.  %Section \ref{UBSR} discusses our universal syntactic representation (UBSR) and section \ref{higher-order} studies the higher-order concepts layered over the UBSR respectively. Section \ref{sys-arch} explores our profiler design while section \ref{impl} examines implementation details. Section \ref{exp} evaluates our experimental findings. 
% \vspace{-5pt}

\section{Related Work}
\label{refwork}
\textit{Data Profiling}: Data profilers \cite{metareader, raha2019, epperson2023deadalivecontinuousdata, huang2024cocoonsemantictableprofiling} in tabular/structured data settings play an important role in the understanding data sets for aiding data curation lifecycle \cite{10.1145/3329486.3329499,47967} by identifying data issues such as missing, extreme,
or erroneous values. Structured metadata extraction from corpora in unstructured data domains such as web pages \cite{snowball2000}, images \cite{shredder12} and pdf documents \cite{lin2024accurateefficientdocumentanalytics} have been explored in prior works. In contrast, in this work we focus on data profiling of code data. 

\textit{Code Parsers / Multi-Lingual Code Analyzers: }While lexical analyzers and parsing tools such as Tree-Sitter \cite{treesitter} and Antlr\cite{antlr} support multiple languages, the structure extracted by these parsers are rooted in language-specific grammars and requires post-processing to support unified language-agnostic data profiling. Open source projects such as Babelfish \cite{babel} and Kythe \cite{kythe} have been proposed as universal code schemas with support for a limited number of languages. In contrast, our tool targets use in production-grade LLM data curation settings supporting over 200+ languages. Further, the code sample representation used in these tools is not tabular, which is a key requirement for interactive "queryable" data profiling.  

\textit{LLMs For Entity Extraction From Unstructured Data}: Data cleaning in the machine learning context is a well-studied topic \cite{côté2024datacleaningmachinelearning}. The use of LLMs for data curation has been proposed both in the structured setting and the unstructured setting. In the structured setting, LLM-based approaches for data curation such as \cite{LLMStructCuration1, LLMStructCuration2, LLMStructCuration3} have been proposed for curation tasks such as entity matching \cite{papadakis2020surveyblockingfilteringtechniques}, error detection \cite{rekatsinas2017holocleanholisticdatarepairs,holisticcleaning} and data imputation \cite{Liu_2023} extracting structure from unstructured documents. For unstructured data, LLMs have been proposed for extracting structured views from unstructured data either by generating entity extraction code \cite{arora2023languagemodelsenablesimple} or directly by extracting entities using prompting \cite{wu2024learningextractstructuredentities}. Recent work \cite{chen2024seeddomainspecificdatacuration} has proposed domain-specific LLM-driven data curation approaches that identify the right mix of code synthesis, direct prompting, vector lookups and use of smaller models on the data curation path to trade-off quality with cost. In contrast to these approaches, we differ along two dimensions. Our hybrid approach of combining a rule-based deterministic online path coupled with a cost-sensitive LLM-based offline phase provides a good balance between quality, generality and cost. Further, unlike text data, our profiler specifically focuses on code data sets where there is a rich underlying lexical structure to the data.

\textit{Syntax-Aware Code Processing For LLM Data Curation}: In contrast to pretraining approaches that are structure-unaware \cite{feng2020codebertpretrainedmodelprogramming,lachaux2020unsupervisedtranslationprogramminglanguages, ahmad2021unifiedpretrainingprogramunderstanding, li2023starcodersourceyou, rozière2024codellamaopenfoundation}, structure-aware approaches have been proposed in deep learning \cite{yin2017syntacticneuralmodelgeneralpurpose, kim2021codepredictionfeedingtrees, articletreegen} and more recently in LLMs \cite{roziere2021dobfdeobfuscationpretrainingobjective, zügner2021languageagnosticrepresentationlearningsource, wang-etal-2021-codet5, guo2021graphcodebertpretrainingcoderepresentations, gong2024astt5structureawarepretrainingcode, tipirneni2024structcoderstructureawaretransformercode}. Similarly, structure-aware approaches have also been proposed for fine-tuning code models \cite{wu2024structureawarefinetuningcodepretrained, tsai2024codelessalignmore} and for code RAG \cite{poesia2022synchromeshreliablecodegeneration,du2024codegragextractingcomposedsyntax}. Our syntactic profiler is complementary to these techniques as it can be used to universalize syntactic concepts from a large set of languages, thus serving as the basis for common tasks such as multi-lingual data processing in the context of syntax-aware model training and for code search.

% \vspace{-5pt}
\section{Solution Outline}
\label{sol-outline}
In this section, we discuss the key requirements of a multi-lingual code data profiler, how they motivate our design decisions, and outline our overall approach. 
% \vspace{-8pt}
\subsection{Key Requirements for Multilingual Code Data Profiler}
We outline the key requirements of our code data profiler below.\\
\textbf{R1: Need for a tabular schema to store metadata}\label{R1}: The purpose of profiling is to help users gain a deeper understanding of the code's data. To support any combination of analytics on the profiled data, we propose that the schema for storing metadata should be organized in a structured, tabular format from unstructured code, allowing for SQL-like querying capabilities. \\ %Extracting a structured tabular representation from unstructured code enables flexible SQL-like querying capabilities that can be exposed to the rest of the profiler. Unlike text, programming languages have well-defined and strict grammatical rules. Further, code is organized hierarchically and is tree-structured. Leveraging this structure for extracting code-specific properties is a requirement. \\
\textbf{R2: Need For Language Agnostic Representation}\label{R2}: 
Different programming languages have distinct syntax constructs. To enable multi-language code profiling, it is necessary to extract a unified representation that can encapsulate the various classes of programming languages. \\ %While the grammatical structure of code makes it more predictable than text, grammatical rules vary across programming languages i.e. while programming concepts across languages are similar, they are not identical. Thus, the ability to extract a unified representation across multi-lingual code data is becomes important. \\
\textbf{R3: Cost-Sensitive Design}\label{R3}: Any data profiling solution must scale to large data set sizes. While it is tempting to leverage the superior pattern-matching capabilities of LLM for extracting structure from unstructured code data, solutions that indiscriminately apply LLMs to every sample on the profiling data path are expensive and inelastic. So we need parsimonious approaches that use precious LLM GPU resources judiciously. \\
\textbf{R4: Need For Syntactic and Semantic Concept Customization}\label{R4}: Settings such as fine-tuning rely on customization for targeted use cases. Thus the ability to experimentally identify customized syntactic and semantic concepts for partitioning code samples becomes essential to comprehend code data properties. We thus need the ability to allow users to customize these syntactic and semantic dimensions to match use cases.

\begin{figure}[h]
\centering
%\vspace{-10pt}
\includegraphics[width=0.8\linewidth]{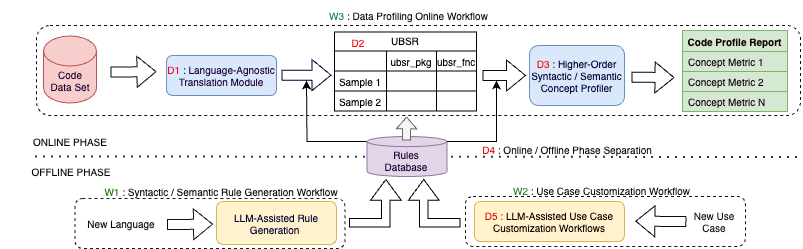} 
%\vspace{-10pt}
\caption{High Level System Design.}
\label{fig:sys-overview}
\end{figure}

% \vspace{-10pt}
\subsection{Design Discussion for Multilingual Code Data Profiler}
Following the above requirements, we next discuss the high level design for our system in this subsection represented by Figure \ref{fig:sys-overview}. For each new target language, the \textit{offline phase} is used to generate deterministic rules leveraging knowledge of LLMs using exemplar code samples from target language. Workflow W1 allows us to generate rules on syntactic constructs from exemplar code samples, and W2 allows us to define semantic dimensions for profiling. Next, we extract the rules that map syntactic concepts to the above-defined semantic concepts. All these rules are then populated in the rule database to be leveraged in the online phase. The online phase is used to generate profiling results on the fly for any new code snippets. This is done by extraction of concepts from the code snippets using the rules in the database and storing them in a tabular schema. The tabular schema enables deriving higher order concept columns, all of which are used to generate profiling reports.  % and then generate rules generate rules on semantic dimensions (eg xyz) and concept tags within each dimension. Next, in the offline phase, we extract rules that map syntactic concepts to the above-defined semantic concepts, and then again populating those rules into the rules database (covered as part of workflow W1). 
 %In the \textit{online phase} outlined in W3 in Figure \ref{fig:sys-overview}, for each new data set, the rules database is consulted for looking up relevant rules to extract concepts from each code sample into the tabular form, augment these with the derived higher-order concept columns and use this to generate code data profiling reports.  %first uses an LLM to generate rules for extracting each universal concept type e.g. packages, and populates these into the rule database (W1 in Figure \ref{fig:sys-overview}). For each new use-case, we support workflows for the definition of use-case customized semantic dimensions and concept tags within each dimension (W2 in Figure \ref{fig:sys-overview}). \footnote{\scriptsize{Semantic dimensions and concept can be added independently of languages and are selected based on use case requirements.}} After this, LLMs are again used to extract rules that map syntactic concepts to the above-defined semantic concepts, again populating those rules into the rules database (covered as part of workflow W1). In the \textit{online phase} outlined in W3 in Figure \ref{fig:sys-overview}, for each new data set, the rules database is consulted for looking up relevant rules to extract concepts from each code sample into the tabular form, augment these with the derived higher-order concept columns and use this to generate code data profiling reports. \\
 We next describe our design choices that enable this workflow and also serve the requirements mentioned above. Detailed design discussion follows in Section \ref{sysarchsec}.\\
\textbf{D1: Tabular Schema For Flexible Querying}: Due to the flexibility provided by structured representations for performing analytics, we use a tabular schema based on the code's abstract syntax tree (AST) to meet requirement R1. AST features provide detailed, structured representations of code samples. We organize each code-AST sample into a consistent tabular format, allowing SQL-like queries using tools like Pandas. This makes it easier to create more complex concepts from the tabular data. \\ %Given the  flexibility of structured representations for both user understanding and downstream analysis, we choose a tabular schema based on the code abstract syntax tree (AST) for addressing requirement R1. AST features serve as fine-grained structured representations of code samples. We marshal each code-AST sample into a size-invariant tabular schema, enabling SQL-like querying over standard data-parallel interfaces such as Pandas. This also facilitates composability of higher-order concept columns from the tabular schema.  \\ 
\textbf{D2: Unified Base Syntactic Representation (UBSR)}: We define a novel UBSR that serves as a common base representation across varied programming languages to address R2. An example of this is $ubsr$-$pkg$ which is a consistent way of representing packages across languages.  \\ %Rather than use per-language tabular representations using ASTs, we instead universalize the tabular schema used to represent multi-lingual code samples for addressing requirement R2. This serves as a common language-agnostic tabular schema for representing multi-lingual code samples.  \\
\textbf{D3: Higher-Order Syntactic and Semantic Concepts}: D2 enables us to have common programming constructs across languages represented in a uniform way. These can then be combined to form higher-order syntactic and semantic concepts that are derived from the base concepts. We think this is important as it gives a user the flexibility to also define custom concepts guided by their use cases, as discussed in R4. An example of a higher order concept is the application domain which the code sample maps to that can be derived from packages, which is a base concept. \\
%Having a set of unified language-agnostic base concepts serves as basis for mapping / composing higher-order syntactic concepts(e.g. code comment ratio) and semantic concepts as per R4 (e.g. the application domain of code snippet) \footnote{\scriptsize{A higher-order concept is a concept derived from one or more concept fields in the UBSR}}. This aids in use case customization of code data profiling. Layering custom concepts over the UBSR (D2) enables high-order concept reuse across languages. \\
\textbf{D4: Online vs Offline Phase Separation}: We design our system to work in two phases: offline phase and online phase. The offline phase uses LLMs to derive rules per programming language, and the online phase uses these rules to profile any new code file. This hybrid approach allows us to use expensive GPUs optimally, in service of R3. \\% While our approach leverages the power of LLMs to learn rules over unknown languages from exemplary code samples derived from a small set of ``known'' languages, we do not use LLMs on our data path (R3 \ref{R3}). Instead, we use LLMs in an offline phase to learn these rules. Our online phase (executed per sample) avoids GPUs entirely. \\
\textbf{D5: Workflow For Identifying Semantic Dimensions and Concepts}: Our semantic profiler uses an LLM-integrated workflow for defining (and refining) use-case customized dimensions and concepts for semantically partitioning data, as needed by R4. %Our LLM workflows help in this use-case customization process (R4 \ref{R4}), helping the user extract insights from code data at user-defined concept granularity. 

%\vspace{-7pt}
%\subsection{Profiler Workflow}
%Now, that we have discussed the design choices The key workflows for the data profiler are also outlined in Figure \ref{fig:sys-overview}. For each new target language, the \textit{offline phase} first uses an LLM to generate rules for extracting each universal concept type e.g. packages, and populates these into the rule database (W1 in Figure \ref{fig:sys-overview}). For each new use-case, we support workflows for the definition of use-case customized semantic dimensions and class tags within each dimension (W2 in Figure \ref{fig:sys-overview}). \footnote{\scriptsize{Semantic dimensions and classes can be added independently of languages and are selected based on use case requirements.}} After this, LLMs are again used to extract rules that map syntactic concepts to the above-defined semantic classes, again populating those rules into the rules database (covered as part of workflow W1). In the \textit{online phase} outlined in W3 in Figure \ref{fig:sys-overview}, for each new data set, the rules database is consulted for looking up relevant rules to extract concepts from each code sample into the tabular form, augment these with the derived higher-order concept columns and use this to generate code data profiling reports.

\section{System Architecture}
\label{sysarchsec}
In this section, we discuss a code profiling framework that realizes our design goals. Section 4.1 proposes Unified Base Syntactic Representation as a means of addressing D1 and D2. Section 4.2 discusses our approach to support customizable higher-order concepts (D4). Section 4.3 examines the cost-sensitive decomposition of the problem into online and offline phases (D3). Section 4.4 discusses our use-case customization workflows for aiding the definition of semantic dimensions and concepts.

\subsection{Unified Base Syntactic Representation}
\label{sec:UBSR}

Our proposed Unified Base Syntactic Representation (UBSR) is a standardized tabular schema designed to abstract programming language-specific details from unstructured multilingual code data. This abstraction helps data engineers and scientists better understand the characteristics of code data for selection and curation, while also providing a common schema for language-agnostic downstream code analysis. The representation aims to capture shared features across different programming paradigms. As a unified structured framework, it serves as a foundation for extracting higher-order syntactic and semantic properties, enabling rich, use-case-specific partitioning of code data as discussed in Section \ref{sec:higher-order}.

\subsubsection{Base Syntactic Concepts Across Languages}
While programming languages expose similar syntactic building blocks to represent programming intent, such as importing packages/libraries, functions, classes, loops, conditionals, comments and others, these concepts are expressed through language-specific grammar, defined by distinct keywords and syntactic form. These syntactic blocks may represent a common concept functionally, but their grammar can vary both within and across languages, making concept extraction grammar-specific and non-trivial. Our framework abstracts language-specific concepts by translating them to \textit{unified base syntactic concepts}, which are encoded as UBSR schema fields. While our set of base syntactic concepts is extensible, we focus here on packages, functions, and comments that are specifically useful in data profiling : (a) \textbf{Packages}: Package names indicate what the code does (e.g. presence of scikit-learn packages may indicate that the code is related to functional category "machine learning"), which framework the code is related to (e.g. Spring vs Hibernate) etc. Thus, a rich set of higher-order semantic concepts can be derived from package names. (b) \textbf{Comments}: The size and frequency of comments in code is the basis for higher-order concepts such as code-comment-ratio that is used as a signal for data quality. 
(c) \textbf{Functions}:  Functions are key to understanding the logical building blocks within a codebase and aid in defining several metrics for profiling, such as average-function-length for characterizing code modularity and thereby to score the quality of code. Function names also help characterize semantic intent.  
%(d) \textbf{Others}: While additional syntactic constructs are valuable for complete syntactic understanding and static code analysis, they contribute less to the curation process. Therefore, our focus remained on extracting package, function, and comment concepts.

%These syntactic concepts allow the UBSR framework to provide a detailed and structured representation of code, enabling more effective analysis across different languages. By extracting these concepts uniformly, the UBSR facilitates scalable code profiling across multiple languages, eliminating the need for language-specific parsers. 

\subsubsection{Representation}
\label{subsec:representation}

\begin{figure}[h]
    \centering
    \includegraphics[width=0.65\linewidth]{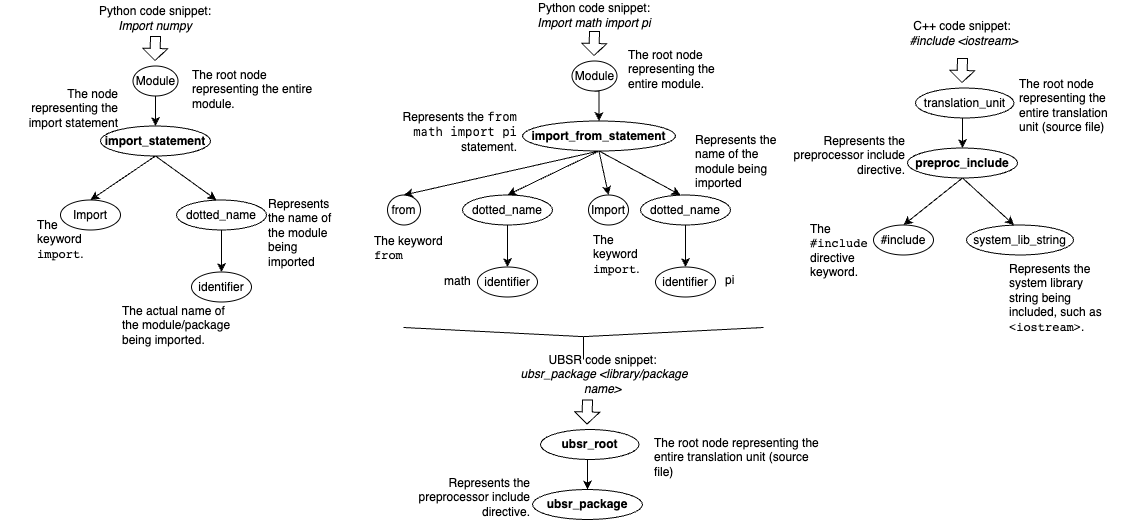}
    %\vspace{-10pt}
    \caption{AST-based Syntactic Variants that Represent the Same Concept.}
    \label{fig:AST-ex}
\end{figure}

The foundation of the UBSR framework is the language-specific AST (Abstract
Syntax Tree) based representation of code, which provides a structured, hierarchical view of the code, enabling precise extraction of syntactic constructs across various programming languages \cite{treesitter}. Consequently, analogous syntactic concepts may be represented by different node types within and across languages, making concept extraction highly dependent on a language’s grammar and rendering the code profiling implementation non-generalizable. For instance, a Python \texttt{import\_statement} and \texttt{import\_from\_statement}, and a C++ \texttt{\#include} directive, all serve similar purposes by incorporating external libraries or packages into the code. However, they may be represented by distinct node types in their respective ASTs as shown in Figure \ref{fig:AST-ex}. This is unsurprising, as ASTs are constructed based on well-defined language-specific lexical rules, which differ from one language to another. 

To reduce the complexity, we take two steps: group languages based on paradigms and define new universal nodes to capture base concepts across languages. We group languages into three primary paradigms that enable us to effectively extract syntactic constructs while minimizing the complexity of language-specific variations inspired by \cite{scott2016programming}. These paradigms are (a) \textit{C-like Syntax:}  Imperative, procedural languages exhibit syntax similar to C, including braces for code blocks and semicolons to end statements. (b) \textit{Scripting and Dynamic Syntax:} Flexible languages that support dynamic typing and features like first-class functions and dynamic objects, enabling concise and readable code. (c) \textit{Functional and Expression-Oriented Syntax:} Using functions as primary building blocks, these languages support higher-order functions, immutability, and expression-based constructs.

Next, we define new universal nodes to consistently represent syntactic concepts across languages: \texttt{ubsr\_package} for import functionalities,  \texttt{ubsr\_comment} for \texttt{comments} and  \texttt{ubsr\_function} for \texttt{function}.
%by mapping language-specific library/import node types to the syntactic concept node of type \texttt{ubsr\_package}. Additionally, the UBSR framework maps AST nodes representing \texttt{comments} to \texttt{ubsr\_comment} and \texttt{function} to \texttt{ubsr\_function} in the UBSR. 
To maintain metadata about the code snippet, such as the programming language, we introduce a root node, \texttt{ubsr\_root}. As shown in figure \ref{fig:AST-ex}, \texttt{ubsr\_root} is a parent of node \texttt{ubsr\_package}. The hierarchical nature of the AST is preserved in the UBSR through a field called edges that denote parent-child relationships between nodes. These relationships are mapped to UBSR fields. We also add metadata based on node relationships, which enables semantic analysis like data dependencies. It's worth noting that collapsing the code corresponding to child nodes into concept nodes simplifies and normalizes the representation, making it easier to annotate the entire subtree within a single concept node (see figure \ref{fig:AST-ex} for example). This collapsing technique reduces representation complexity while retaining the essential syntactic information needed for consistent analysis across different programming languages.

\begin{table}[h]
\centering
\scriptsize
\caption{UBSR Schema Representation.}
%\vspace{-10pt}
\label{tab-ubsr-schema}
\begin{tabular}{|p{3cm}|p{3.7cm}|p{6.2cm}|}
\hline
\textbf{Key} & \textbf{Possible Values} & \textbf{Description} \\ \hline

\multicolumn{3}{|l|}{\textbf{\texttt{"nodes":}}} \\ \hline
\texttt{"id"}        & Integer (e.g., \texttt{0}, \texttt{1})     & Unique identifier of the node. \\ \hline
\texttt{"code\_snippet"} & String (e.g., \texttt{"ubsr\_package math"}) & A snippet of code or a description of the node. \\ \hline
\texttt{"node\_type"} & String (e.g., \texttt{"ubsr\_root"}, \texttt{"ubsr\_package"}, etc.) & Type of node representing various syntactic concepts. \\ \hline
\texttt{"parents"}   & Array of Integers (e.g., \texttt{[1, 2]}) & List of parent node IDs. \\ \hline
\texttt{"children"}  & Array of Integers (e.g., \texttt{[1, 2]}) & List of child node IDs. \\ \hline
\multicolumn{3}{|l|}{\textbf{\texttt{"metadata"} (within nodes):}} \\ \hline
\texttt{"info"}      & String & General information about the node. \\ \hline
\texttt{"language"}  & String (\texttt{"cpp"}, \texttt{"python"}, etc) & Programming language of the node. \\ \hline
\texttt{"original\_code"} & String (e.g., \texttt{"int main() {...}"}) & Original code snippet corresponding to the node. \\ \hline
\texttt{"loc\_original\_code"} & Integer & Line of code of the concept \\ \hline

\multicolumn{3}{|l|}{\textbf{\texttt{"edges":}}} \\ \hline
\texttt{"directed\_relation"} & String (\texttt{"parent\_node"}) & Type of relationship between nodes e.g. parent-child. \\ \hline
\texttt{"metadata"}  & Object & Additional metadata for the edge, which can be empty. \\ \hline

\end{tabular}
\end{table}

The process of converting the language-specific ASTs into UBSR form involves recursively parsing the hierarchical AST and mapping AST node and edge types into UBSR fields. The \texttt{metadata} fields—custom information string, programming language, and code\_snippet enhance the UBSR's usability for downstream tasks. The custom information string allows for the addition of context-specific data, making the representation adaptable. The language tag ensures accurate handling of multi-language codebases. The \texttt{code\_snippet} stores the collapsed code under the nodes. UBSR representations of each data point are stored in a tabular format, organizing nodes and edges as columns of a wide table. As discussed earlier, such a tabular representation enables easy data-parallel querying by downstream modules via SQL/Spark style interfaces. Table \ref{tab-ubsr-schema} describes the fields of the UBSR which maps AST nodes to a well-defined schema. The schema captures both syntactic nodes (derived from AST node types) and relationships between syntactic nodes (derived from AST edges).  %The UBSR thus serves as the lowest level and finest-grained structured representation for multi-lingual code data.
%\subsubsection{Leveraging Syntactic Paradigms for Unified Concept Extraction}
%Syntactic paradigms define the styles and structures of programming languages, influencing their syntax, semantics, and design \cite{scott2016programming, sebesta2015concepts}. By grouping languages into the following three primary paradigms, we effectively extract syntactic constructs while minimizing the complexity of language-specific variations (a) C-like Syntax:  Imperative, procedural languages exhibit syntax similar to C, including braces for code blocks and semicolons to end statements. (b) Scripting and Dynamic Syntax: Flexible languages that support dynamic typing and features like first-class functions and dynamic objects, enabling concise and readable code. (c) Functional and Expression-Oriented Syntax: Using functions as primary building blocks, these languages support higher-order functions, immutability, and expression-based constructs.
Our UBSR framework currently supports 21 languages across the above-mentioned syntactic paradigms. Our UBSR framework unifies code representations by categorising languages based on these syntactic paradigms as shown in Table \ref{ubsr-support}. In the table, \texttt{NA} denotes concepts not present in the language. %, while asterisks (*) mark languages whose support was manually handcrafted into the framework and serves as context (prompt examples) for deriving rules for other languages.

\begin{table}[h]
\centering
\scriptsize
\caption{Base Syntactic Concepts Supported by the UBSR across Different Syntactical Paradigms.}
%\vspace{-10pt}
\label{ubsr-support}
\begin{tabular}{|p{3.4cm}|p{5.2cm}|p{0.9cm}|p{0.9cm}|p{1.1cm}|}
\hline
\textbf{Syntactical Paradigms} & \textbf{Languages supported (Known*)} & \textbf{Package} & \textbf{Function} & \textbf{Comment} \\ \hline

\multirow{1}{*}{\textbf{C-like Syntax}} 
& \textbf{C}*, \textbf{Java}*, \textbf{C}\#, \textbf{CPP}, \textbf{Objective C}, \textbf{Rust}, \textbf{Golang}, Kotlin             & Yes  & Yes & Yes \\ \cline{2-5}
\hline

% \iffalse
% & \textbf{Java}*       & Yes  & Yes & Yes \\ \cline{2-5}
% & \textbf{C}\#         & Yes  & Yes & Yes \\ \cline{2-5}
% & \textbf{CPP}         & Yes  & Yes & Yes \\ \cline{2-5}
% & \textbf{Objective C} & Yes  & Yes & Yes \\ \cline{2-5}
% & \textbf{Rust}        & Yes  & Yes & Yes \\ \cline{2-5}
% & \textbf{Golang}      & Yes & Yes & Yes \\ \cline{2-5}
% & Kotlin      & Yes  & Yes & Yes \\ \hline
% \fi

\multirow{2}{*}{\makecell{\textbf{Scripting and Dynamic}\\ \textbf{Syntax}}} & \textbf{Python}*, \textbf{JavaScript}*, \textbf{Dart}, \textbf{Typescript} & Yes  & Yes & Yes \\ \cline{2-5}
& QML         & Yes  & NA & Yes \\ \cline{2-5}
& \textbf{Perl}        & Yes & Yes & NA \\ \hline

% \iffalse
% & \textbf{JavaScript}* & Yes  & Yes & Yes \\ \cline{2-5}
% & \textbf{Dart}        & Yes  & Yes & Yes \\ \cline{2-5}
% & QML                  & Yes  & NA & Yes \\ \cline{2-5}
% & \textbf{Typescript}  & Yes  & Yes & Yes \\ \cline{2-5}
% & \textbf{Perl}        & Yes & Yes & NA \\ \hline
% \fi

\multirow{2}{*}{\makecell{\textbf{Functional and}\\ \textbf{Expression-Oriented} \textbf{Syntax}}} 
& \textbf{Haskell}*, Elm*, Agda, \textbf{D}, \textbf{Nim}, \textbf{Scala}      & Yes  & Yes & Yes \\ \cline{2-5}
& \textbf{Ocaml}       & Yes  & NA & Yes \\ \cline{2-5} \hline

% \iffalse
% & Elm*        & Yes  & Yes & Yes \\ \cline{2-5}
% & Agda        & Yes  & Yes & Yes \\ \cline{2-5}
% & \textbf{D}           & Yes  & Yes & Yes \\ \cline{2-5}
% & \textbf{Nim}         & Yes  & Yes & Yes \\ \cline{2-5}
% & \textbf{Ocaml}       & Yes  & NA & Yes \\ \cline{2-5}
% & \textbf{Scala}       & Yes  & Yes & Yes \\ \hline
% \fi

\end{tabular}
\end{table}

\subsection{Higher-Order Syntactic and Semantic Concepts}
\label{sec:higher-order}
Our code data profiler utilizes the base syntactic concepts in the code sample UBSR to compose higher-order syntactic as well as complex semantic concepts as discussed below.

\subsubsection{Higher Order Syntactic Concepts} The base concepts in the UBSR can be used to derive higher-order concepts related to the syntactic structure of code blocks that are used for partitioning and analyzing code data properties. For example, 
% the count of control-flow decision node types in the UBSR (if, while, for node types) can be used to derive the number of independent paths in a program, which in turn characterizes the \textit{cyclomatic complexity} of the code block. Similarly 
\textit{code comment ratio} (CCR), which is computed by dividing the total lines of code by total comment lines, can serve as a useful profiling metric. We call such concepts higher-order syntactic concepts since they are derived from base UBSR syntactic concepts\footnote{Other examples of higher-order syntactic concepts relevant to profiling include \textit{cyclomatic complexity}, \textit{mean nesting depth}, and \textit{mean function fan-in, mean function fan-out} (number of invocations from and into a function).}. In the context of data profiling, the ability to characterize code in terms of such higher-order syntactic concepts 
% and specifically to quantitatively characterize them for thresholding or scoring purposes 
serves as the basis for curating code data points.
% and deciding which data to retain and which to discard. 
For example, 
% we may decide that we want to discard 
data points with low CCR may be preferred over those with high CCR.
% is less than a user-provided threshold. 
Our code data profiler gives users the ability to define proprietary and complex higher-order syntactic concepts specific to their requirements using UBSR base concepts using rules as discussed in Section \ref{sec:sem_imp}.

\subsubsection{Semantic Concepts}\label{sem-con} The idea of semantic profiling of a code dataset is to derive semantic attributes for input code snippets as a complementary profiling information to the syntactic concepts. % from UBSR fields in a given dataset that help a user map to their end usecases. Such \textit{semantic dimensions}, which are use-case customizable, include the data partitioning criteria such as functionality, end application domain that the code is meant to serve, or programming framework(s) that the code conforms to. 
We use the term ``Semantic dimension" to represent a pre-defined semantic concept which is used as a dataset partitioning criteria and is re-usable across many use-cases. We choose the following semantic dimensions and concepts in our system: Semantic dimension: "Functionality"; Concepts: "GUI Design", "Networking and Communication", and "Mathematics", etc. Semantic dimension: "Application Domain"; Concepts: "Mobile Development", "Game Design", "Internet of Things (IoT)", etc. Semantic dimension "Coding Frameworks"; Concepts: "Django", "REST", "SpringBoot", etc. All the semantic concepts are finally used as profiling metrics and shown to the end user. These are derived from UBSR using LLM aided method that we detail in Section \ref{on-off-sem}, and can be customised as per user's needs. 

\subsection{Offline and Online Phase Separation}\label{sys-arch}
\begin{figure}[h]
\centering
\includegraphics[width=0.8\linewidth]{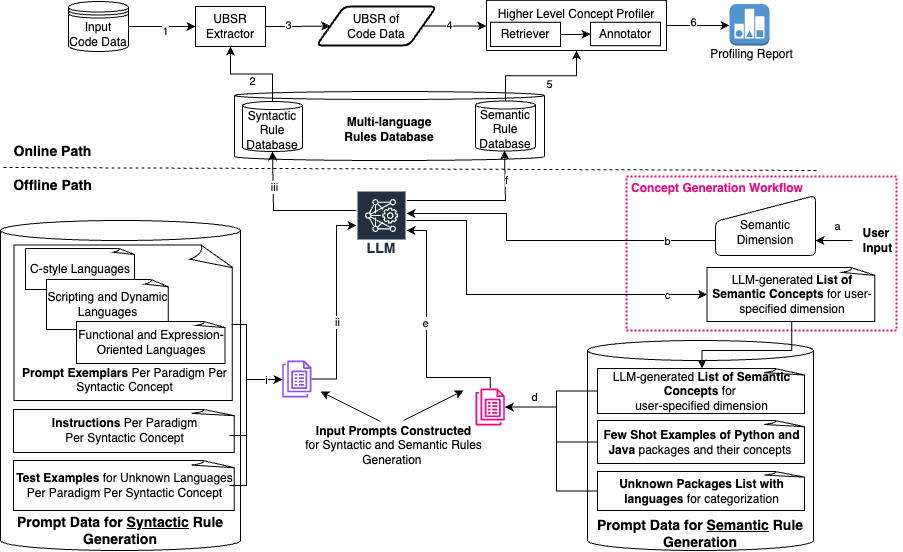}
%\vspace{-8pt}
\caption{End-to-end Workflow of the Proposed Code Data Profiling Framework.}
\label{fig:e2e-wf-ubsr}
\end{figure}

The system architecture of the proposed code data profiling framework, as shown in Figure \ref{fig:e2e-wf-ubsr}, is designed to accurately extract base syntactic and higher-order syntactic/semantic concepts from unstructured code data, enabling language-independent code profiling. The \textit{online path} involves a sequence of steps undertaken for translating each unstructured code data point into its corresponding UBSR (captured in steps 1, 2, and 3), from which higher-order syntactic and semantic concepts are extracted (captured in steps 4, 5, and 6). In contrast, the LLM-aided \textit{offline path} is responsible for generating language-specific rule sets by prompting an LLM and populating them into the multi-language base syntactic rule database (steps i, ii, and iii), and into the semantic rules database (steps a - f) as shown in Figure \ref{fig:e2e-wf-ubsr}. The paths are discussed in detail as follows:

% . The online path is responsible for translating unstructured code data into the UBSR using deterministic language-specific base syntactic rules that have been pre-populated (by the offline path) into the multi-language rules database, as illustrated in steps 1, 2, and 3 of Figure \ref{fig:e2e-wf-ubsr}. The generated UBSR is then used by the downstream profiler modules to extract higher-order syntactic and semantic concepts in steps 4, 5, and 6. These higher-order rules are defined by data scientists and domain experts to extract these higher-order concepts from the generated UBSRs. 
%This dual-path approach ensures that the online processing is continuously refined and improved by the insights gained from offline processing, thereby enhancing the overall efficiency and accuracy of the system. 

% It also facilitates rules for higher-order concept extraction, as illustrated in steps a to f. 

\subsubsection{Online and Offline Path for UBSR Generation}
Base syntactic concept extraction from code blocks into UBSR format is the core function of the UBSR generation framework. This process relies on pre-loaded base syntactic rule sets, which are used by the UBSR extractor to accurately map code snippets into the UBSR structure. Each code snippet is then parsed into a language-specific AST, which is recursively traversed to map node and edge types to their corresponding language-agnostic UBSR types while preserving the hierarchical relationships within the code. Below, we describe the core components that enable syntactic concept extraction into UBSR form.

\textit{Extensible Base Syntactic Rule Database}: Each programming language has its way of representing syntactic concepts, which are represented by language-specific node types in the AST. The base syntactic rule database contains a comprehensive collection of deterministic rules that associate the language-specific AST node and edge types to their corresponding UBSR node and edge types. In addition to the AST to UBSR node type mapping, the rule also comprises of \textit{extractor code} that is responsible for parsing the code snippet in the span of the retrieved AST node and extracting the relevant node-type specific attributes (such as package name, function name, comment contents) corresponding to each AST node. 

%Key functions are \textbf{ (a) Standardization Across Languages}: The rule set ensures that, despite the syntactic differences between languages, essential syntactic features are uniformly captured within the UBSR. For example, a package import in Python and the inclusion of a header file in C++, though represented differently in their respective ASTs, would be mapped to a common UBSR node type, such as \texttt{ubsr\_package}. \textbf{(b) Syntactic Desugaring}: Due to language-specific grammar, a single syntactic concept can be represented by different node types in the AST of the same language. For instance, Python uses distinct node types for package imports, such as \texttt{import numpy}, \texttt{import numpy as np}, or \texttt{import numpy, pandas}. This grammatical variation underscores the need for multiple rule variants to map to the same node type. The UBSR, therefore, provides a "desugared" representation of these syntactic features, simplifying and unifying them across languages. \textbf{(c) Extensible and Future-Proof:} As languages evolve or new ones emerge, the extensible rule engine, integrated with LLM capabilities, allows developers to expand the syntactic rule set, mapping new constructs to UBSR node types.

\textit{Offline Path for Base Syntactic Rule Generation}: The offline path is responsible for synthesizing rules that are pre-loaded into the rules database and are leveraged by the online path. This path utilizes the few-shot Chain of Thought prompting technique \cite{wei2023chainofthoughtpromptingelicitsreasoning} to guide the LLM through a step-by-step process for rule generation, using carefully selected prompt exemplars and instructions tailored to the test language's syntactic paradigm and the concept to be extracted. This method ensures that the LLM can accurately generalize from the provided examples to generate effective rules for a variety of languages. We break down the prompt template into various sections and explain how the contents of these sections affect the rule extraction process below.

%\footnote{The detailed prompts for each concept are available in the folder *FSE-25/prompts/ubsr*, which can be accessed through the data shared at \url{https://doi.org/10.5281/zenodo.13757379}.}

\textbf{Prompt exemplars} dataset is organized into three \textit{syntactic paradigms} mentioned above. For each paradigm, we select a set of known programming languages for which base syntactic rules are handcrafted. Exemplars from these languages provide context for the larger set of test languages in that same paradigm. LLMs are prompted using these exemplars. Each exemplar contains minimal code snippets and their corresponding ASTs, designed to capture specific syntactic concepts within each paradigm. This allows the LLM to focus on key AST patterns for rule generation, without being distracted by irrelevant details. Since the extracted rules are deterministic mappings, they can then be applied to arbitrarily complex programs/code blocks in the online path. Selecting few-shot examples from the same paradigms has a \textit{significant} impact on rule correctness. \textbf{Instruction section} outlines the prompt instruction for the LLM, specifying the expected output format and the syntactic concepts to be extracted. The instructions are common across syntactic paradigms for each concept. \textbf{Test input} section includes minimal code snippets that represent the target syntactic concept and their corresponding AST representations. 

\textit{AST Pruning}: Prompt exemplars and test inputs are carefully curated to optimize prompt size, as lower token counts are crucial for LLM efficiency, reducing both processing time and resource usage. The prompt size is also limited by the LLM’s context window, restricting the amount of data it can process at once. For instance, Llama 3 70B has a context window of 8k tokens \cite{meta-llama3-instruct-70b}. To enhance efficiency, we implemented AST pruning to minimize AST size in prompt exemplars and test inputs, allowing the LLM to focus on relevant sections, improving both accuracy and computational efficiency. Depth-level pruning restricts the AST to the most relevant levels, simplifying its structure while preserving key syntactic information. Common concepts like packages and functions are often found at the first level of the AST, allowing for effective pruning of higher-level nodes, while other concepts like comments may reside deeper. To capture these deeper concepts, the AST must be extended, but this can also introduce unnecessary nodes, leading to an increased token count. Concept-level pruning addresses this by targeting only the nodes relevant to the desired syntactic concepts, regardless of depth, streamlining the AST for more efficient rule generation.

\subsubsection{Online and Offline Path for Higher Order Concepts}\label{on-off-sem}
Our higher-order code data profiler is equipped with a novel LLM-aided module that is capable of extracting higher-level syntactic and semantic concepts (henceforth referred to as higher-level concepts) from the base syntactic concepts obtained from the UBSRs of code snippets. As shown in Figure \ref{fig:e2e-wf-ubsr}, the higher-level concept profiler module includes three major components: (a) Retriever, (b) Annotator, and (c) Higher Level Concept Mapping RuleSet, which work together in the online phase to generate higher level concept annotations. The \textit{retriever} retrieves the relevant syntactic information, such as the number of lines of code and comments, names and arguments of functions, list of imported packages, and the programming language, from the UBSR of a code snippet.  The \textit{annotator} determines the values of the higher-level concepts using the extracted syntactic information and annotates the input UBSR dataset with these concepts. For obtaining the higher level concepts associated with the base syntactic concepts, the annotator utilizes the \textit{Higher Level Concept Mapping Rule-Set} which contains the rules for the calculation of the higher level syntactic concepts as well as the mapping of the base syntactic concepts to the most relevant semantic concepts. The specific approach to obtain the semantic mappings is discussed as follows.

\textbf{LLM-Aided Offline Path for Semantic Ruleset Generation:} Given a set of semantic concepts and a set of previously unseen packages, the package-to-concept mapping rules are generated through a LLM prompting technique, discussed as follows. Note that these concepts are customizable based on the user requirement of semantic dimension, the process of which is discussed in Section \ref{sec:use-case-custom}. For rule-set generation, we utilize a prompt, termed as the ``Semantic Mapping Prompt'', with the following information embedded in it: (a) Concept List, and (b) Set of examples containing well-known packages and their concepts. The list of packages to be categorized are given as inputs to the prompt, and the outputs obtained from the LLM are then processed and stored in the Semantic Ruleset Database. Note that, the examples provided in this prompt primarily serve the purpose of teaching the LLM the desired format of output. This stage is repeated for all packages present in the input data set having no entries in the rules database. The design of the prompt and other implementation details are mentioned in Section \ref{sec:sem_imp}.

\textbf{Online Path for Semantic Profiling:} The main idea behind our approach for semantic concept mapping is that the base syntactic concept in the UBSR of a code snippet can help in identifying standardised APIs that have well-defined functionalities and their information is typically available in the public domain. We primarily focus on the programming language packages as a base concept to derive semantic concepts. Our module utilizes the information of the imported packages in a code snippet to understand their functionalities or downstream usecase, thereby enabling the extraction of rich semantic information. The mapping of packages in different programming languages to the most relevant semantic concepts is stored in the semantic ruleset database, which is then used to obtain the semantic mapping of each package. 
% It is noteworthy that, while packages provided a higher level knowledge of the semantics of a code snippet, other syntactic concepts such as functions, comments, and keywords can be used to further refine this knowledge and gain a deeper understanding of the workings of the code. 
Moreover, in our module, the nature and granularity of the semantic characterization of packages can be customized by the user based on the use-case requirements. It also provides the flexibility of combining higher-order syntactic concepts with semantic concepts, which in turn enables data profiling based on complex user-specified criteria. The semantic rules database is populated \textit{apriori} in the offline phase, based on the user-specified semantic dimension through prompting an LLM. In the case where no match is found for a particular package in the semantic rules database in the online phase, the name of the package and its programming language are recorded separately. In a subsequent offline phase, the mapping rules for these recorded packages are generated similarly through LLM prompting and updated in the database. This ensures that the database is customizable and extensible according to the requirements of the downstream usecases. %In the following subsection, we discuss the process of LLM-aided semantic ruleset generation in detail.

% \textcolor{red}{Pls add prompt template}

\subsection{Use Case Customization}\label{sec:use-case-custom}
% The first stage is the one time generation of list of classes for the characterization of packages based on the requirements of the user.
As discussed earlier, our proposed higher-order concept profiler is capable of supporting use-case-driven customization. Shown in the highlighted box (Concept Generation Workflow) in Figure \ref{fig:e2e-wf-ubsr}, the use-case-specific semantic dimensions \ref{sem-con} are taken as an input before profiling. To achieve this, the user may choose a dimension from a pre-defined set or define a new usecase-customized semantic concept as a dimension. This information is then passed to an LLM in a prompt, referred to as ``Concept List Prompt'', to obtain a list of possible concepts of packages as per the given dimension. Iterative prompting is then used to refine the names of the concepts to prevent any overlap between any two concepts as well as to remove any unimportant/niche concept. Finally, the list of relevant, important, and non-overlapping semantic concepts is obtained which is referred to as the ``Concept List". This list serves as one of the inputs to the stage of generation of semantic mapping. The design of this prompt is mentioned in Section \ref{sec:sem_imp}.

\section{Implementation}
\label{impl}
\subsection{Base Syntactic Concept Extraction}
The implementation of the UBSR framework integrates several key components that enable efficient base syntactic concept extraction and dynamic rule generation. We implement all the components of our UBSR framework as Python components. Below, we outline the highlights of our implementation of the online (UBSR translation) and offline (generating new rules using LLMs) phases. The UBSR schema of samples in the code data set are collected in a Pandas dataframe, thereby enabling scalable data-parallel higher-order concept extraction by the higher-order profiler. The dataframe contents can be persisted in multiple formats (Parquet, JSON etc). 

\subsubsection{Extensible Base Syntactic Rule Database}
It is implemented as a flexible structure, using JSON files to map AST node types to UBSR node types, along with extractor functions written in Python. The example rule below is designed to extract the package concept and the package/library name from the AST nodes \texttt{import\_statement} and \texttt{import\_from\_statement} that represent the concept package. Similar rules exist for each base syntactic concept across different programming languages.

{\footnotesize
\begin{lstlisting}[caption={Base Syntactic Rules to Extract a Package Concept from the AST of a Python Code.}]
{
    "import_statement": {
       "ubsr_node_type": "ubsr_package",
       "extractor": "text = code_snippet.split('import')[1].strip() \nif (',' in text):\n    imports = text.split(',')\n    all_imps = []\n    for imp in imports:\n        imp = imp.strip().split(' ')[0].strip()\n        if ('.' in imp):\n            imp = imp.split('.')[0]\n        all_imps.append(imp)\n    all_imps = list(set(all_imps))\n    self.extracted = (', ').join(all_imps)\nelse:\n    imp = text.strip().split(' ')[0].strip()\n    if ('.'in imp):\n imp= imp.split('.')[0]\n self.extracted= imp\n"},
     "import_from_statement": {
       "ubsr_node_type": "ubsr_package",
       "extractor": "text = code_snippet.split('from', 1)[1].strip()\ntext = text.split(' import')[0]\ntext = text.strip()\nif ('.' in text) :\n    self.extracted = text.split('.')[0]\nelse:\n   self.extracted = text\n"}, ...
}
\end{lstlisting}}

%\vspace{-6pt}

\subsubsection{UBSR extractor} We use Tree-sitter, a robust parser that supports over 170 programming languages, to convert the code from various languages to ASTs \cite{treesitter}. %Tree-sitter's versatility allows the UBSR framework to handle multi-language codebases, accurately capturing language-specific syntactic nuances seamlessly. 
UBSR extractor takes code files or snippets as input, converts them into ASTs, and recursively traverses these trees, applying base syntactic rules to map AST nodes to UBSR nodes. The resulting UBSR is then outputted, with nodes and edges stored in a tabular format. After processing a dataset, the UBSRs are stored in Parquet file format for efficient storage and consumption by downstream applications. The pseudocode in Algorithm \ref{pseudo:ubsr-extraction} summarizes the process of syntactic concept extraction within the UBSR framework.

%- Explain a bit about the externalizing the tabular representation in parquet / other forms that you support. 
%Also is the implementation is data-parallel, highlight this aspect. 
%- Mention tree-sitter earlier. Explain and justify why. 
%\vspace{-7pt}
\begin{algorithm}
\scriptsize
\caption{Pseudocode for Syntactic Concept Extraction in the UBSR Framework}
\label{pseudo:ubsr-extraction}
\begin{algorithmic}[1]
\State \textbf{Input:} Source code snippets from the dataset
\State \textbf{Output:} Unified Base Syntactic Representation (UBSR)

\State \textbf{Initialization:}
\State \quad Initialize the language parsers and load the base syntactic rule set
\State \textbf{for} each code snippet in the dataset:
    \State \quad Parse the code snippet into AST and initialize an empty structure for UBSR nodes and edges
    \State \quad RecursiveTraversal(AST\_root, null)
    \State \quad Integrate the generated UBSR nodes into the UBSR structure and store metadata for each UBSR node
\State \textbf{end for}

\State \quad \textbf{Function RecursiveTraversal(AST\_node, UBSR\_node):}
    \State \quad \quad \textbf{if} AST\_node is null \textbf{then}
    \State \quad \quad \quad Return
    \State \quad \quad \textbf{if} rule exists for AST\_node in the base syntactic rule database \textbf{then}  
    \State \quad \quad \quad Map AST\_node type to new UBSR node type
    \State \quad \quad \quad Create a new UBSR node based on the mapping with the extracted \texttt{code\_snippet}, and \texttt{metadata} and add it to UBSR
    % \State \quad \quad \quad Add the new UBSR node to the UBSR structure
    % \vspace{0.25em}
    \State \quad \quad \textbf{if} UBSR\_node is not null \textbf{then}
    \State \quad \quad \quad Add an edge from UBSR\_node to new UBSR node and update their parent-child relationships
    \State \quad \quad \textbf{end if}
    \State \quad \quad \textbf{for} each child node of the current AST\_node
    \State \quad \quad \quad RecursiveTraversal(child\_node, new\_UBSR\_node)
    \State \quad \quad \textbf{end for}
\State \quad \textbf{End Function}

\State \textbf{Output:} Store the nodes and edges in the UBSR structure in the column format
\end{algorithmic}
\end{algorithm}

% \vspace{-5pt}
\subsubsection{Offline Path for Base Syntactic Rule Generation}
A GUI-based tool, developed using the Streamlit framework \cite{streamlit}, serves as the central interface for enabling the Few-shot Chain of Thought (CoT) prompting technique. Our tool uses Python-based \texttt{gen-ai} \cite{genai2024} client APIs to integrate any open source/frontier LLM into the application. This tool allows users to design structured prompts that guide the LLM through the rule-generation process. The interface is designed to be user-friendly, providing various options such as multi-select boxes for choosing Few-shot input languages, selection boxes for input test languages, text input fields for code snippets, and configurable pruning methods (depth level and concept level) for ASTs.

Given a test input, the user can select the prompt exemplars from various paradigms to increase the possibility of generating the correct rule. Additionally, users can select and apply pruning techniques on the ASTs of the prompt exemplars to optimize prompt length. Depth-level pruning is managed by setting depth thresholds in the translation algorithm, ensuring that only the necessary levels of the AST are processed. Concept-level pruning is achieved by tagging nodes with the concept they represent during the initial parsing phase and then filtering out irrelevant nodes before the traversal begins, making it particularly useful for complex languages where syntactic concepts are scattered across multiple levels of the AST.

Few-shot CoT prompting is used within the tool, ensuring that the LLM receives the necessary step-by-step example context to accurately generate base syntactic rules. To the input request, the LLM responds with a rule and the corresponding output when the rule is applied to the test input. If the output is validated against the test input and deemed correct by the user, the tool allows for direct integration of the rule into the base syntactic rule database. This process enables controlled, human-guided expansion and refinement of the rule database, ensuring that accuracy and relevance are maintained as new concepts and languages are introduced.

%By interacting with an API, the tool feeds in carefully selected training exemplars and test inputs, ensuring the prompts are concise and effective. 
%-Rewrite the above sentence, which I have in comments. It is not clear what the API does. Also avoid subjective words like effective (this is for the reviewer to gauge) - stick to objective / disinterested language. 
% The GUI also supports a feedback-driven, iterative refinement process, allowing users to modify instructions based on the feedback of the LLM. 
%  -  Details would be useful on this "iterative refinement process". This is an implementation section, we should be detailed.
% - You have talked in your sys arch section about pruning etc. Where is all that integrated into your implementation. That must be added somewhere.
%- This may be a good place to discuss the recall / precision tradeoff. 
% \vspace{-10pt}
\subsection{Higher Order Concept Profiler}\label{sec:sem_imp}
% \footnote{TO-DO - Rewrite to explain the data-parallel implementation of rule implementation - SQL rules for implementing rules, Pandas UBSR, Joins with Semantic Lists etc. Add a line mentioning approach shared across high-order syntactic and semantic concepts } 
We implement the different components of the online phase of the profiler using Python, and the semantic database is implemented as a data-parallel Pandas dataframe with the columns - \textit{Library Name}, \textit{Language}, and \textit{Concept-<Dimension>}. There are multiple ``Concept-<Dimension>'' columns in the table, one for each of the dimensions specified by the users. The retriever takes as input the UBSR dataframe produced by the UBSR generator, stored in Parquet / JSON format. Thereafter, it retrieves the various relevant syntactic concepts required by the downstream profiler. For higher-order syntactic concepts, complex queries can be written on top of the input dataframe combining multiple rows/columns. The use of Pandas dataframes for representating both UBSR and higher-order concepts ensures that the query processing is scalable and data-parallel across all stages of the profiler. For example, in our implementation, we used the following query to generate the code-to-comment ratio:

% \lstset{
%     basicstyle=\footnotesize\ttfamily,
%     keywordstyle=\color{blue}\bfseries,
%     identifierstyle=\color{red},
%     commentstyle=\color{green!60!black},
%     stringstyle=\color{orange},
%     morekeywords={import_statement, ubsr_node_type, extractor},
%     breaklines=true,
%     columns=fullflexible,
% }

{\footnotesize
\lstset{basicstyle=\ttfamily\color{blue}, breaklines=true, identifierstyle=\color{blue}, frame=single, xleftmargin=3em, xrightmargin=3em}
\begin{lstlisting}
loc_snippet = GET metadata "loc_snippet" FROM root_node DEFAULT 0
total_comment_loc = SUM(GET metadata "loc_original_code" FROM child 
                        IF child.type IS "ubsr_comment" 
                        FOR child IN root_node.children)
CCR = loc_snippet / total_comment_loc IF total_comment_loc > 0 ELSE 0
\end{lstlisting}}
%\vspace{-3pt}

For semantic profiling, the retriever extracts the list of packages per code snippet along with their programming language. These together with the determined higher-level syntactic concepts are then fed into the annotator for further analysis. On receiving these inputs, the annotator implements a Trie data structure to efficiently search the Semantic Ruleset Database and obtain the mapping of the packages. The obtained semantic concepts per code snippet are then added in the form of a list into a new column which is then appended to the input dataset along with the higher-order syntactic concepts. Thereby, the higher-level concept profiler enhances the queriability of the unstructured code dataset even further. On the other hand, the implementation of the offline semantic ruleset generator primarily consists of the Concept List and Semantic Mapping prompts, and the code to interact with an LLM using its corresponding APIs. The packages are input to the LLM in batches depending on the max token generation length supported by the LLM. The design of the two prompts\footnote{The detailed prompts used are present in the folder FSE-25/prompts/semantic-profiling of the data shared at \url{https://doi.org/10.5281/zenodo.13757379}.} is discussed in the following subsections.

\textbf{Design of Concept List Prompt: }The objective of this prompt is to obtain a list of non-overlapping concepts of programming language packages, given a dimension for categorization. We designed the prompt with the following contents: (i) System Instructions: We use the following two variants of system instructions to steer the behaviour of the LLM \textemdash~\textit{You are an enterprise software professional} and \textit{You are a taxonomist for programming language packages}. These resulted in slightly different outputs, with the latter tending towards more comprehensive concept names. (ii) Task: We define the task as follows -- \textit{Your task is to provide a comprehensive, non-overlapping, and flat list of software library concepts based on <Dimension>}. For our experiments, we used the dimension of ``top-level functionality''. Other possible dimensions include frameworks and application domains. (iii) Context: The list of concepts provided by the user which are mandatory to be included for profiling are passed as context to ensure that the output list includes those concepts as well as other non-overlapping concepts. 
% The specific prompt template is as follows.

% {\scriptsize \fbox{\parbox{0.96\textwidth}{\texttt{\textcolor{blue}{You are an enterprise software professional/you are a taxonomist for programming language packages.} Your task is to provide a comprehensive, non-overlapping, and flat list of software library concepts based on top-level functionality. \textcolor{darkgreen}{The output list must include the following concepts: Web development, Security}}}}}\\

On obtaining the list of concepts from the LLM, we perform manual verification and filtering of the concept names to extract a single, comprehensive, and meaningful concept list with no overlaps. This step however is optional. We then use another prompt with the following task to verify if any important concept is missing -- \textit{List all <Dimension>-based concepts of software libraries which are missing in this list and have no overlap with any of the items in this list}, followed by the previously created concepts list. The output of this prompt is used as the final concept list.

\textbf{Design of Semantic Mapping Prompt: }The objective of this prompt is to obtain the most relevant semantic concepts for a set of programming language packages, given the list of concepts as input. We designed this prompt with the following components: (i) System Instructions: In this case, we use the following system instruction -- \textit{You are a discriminating and conservative programming specialist, responsible for classifying programming language packages}. The terms \textit{discriminating} and \textit{conservative} are used in order to ensure that the LLM chooses the most relevant concept for each package and in case none of the concepts are relevant, the LLM does not choose any irrelevant concept. (ii) Task: We describe the task as follows -- \textit{Your task is to categorize the following packages in the given programming languages based on their <dimension>}. This dimension is same as the one provided by the user, used in the previous prompt. (iii) Contextual information: To ground the output of the LLM, we provide the list of concepts to choose from as the context along with instructions -- \textit{Choose the concepts from the following list: <Concept List>. Given the package name and language in tabular format, add a "Concept" column and output the updated tabular data. Do not include concepts outside of this provided list. If you are absolutely not able to categorize a package, categorize it as "Others". Add <end> at the end of your response.} (iv) Few Shot Examples: As mentioned earlier, we also provide a few examples of packages, their programming languages, and their concepts as per the provided concept list in a tabular format. We ensure that the examples contain a good mix of packages belonging to each language and semantic concept.

The LLM outputs the semantic concept mappings for the given set of packages as input, which are then processed and stored in the Semantic Ruleset Database.

% \vspace{-10pt}
\section{Experiment Results and Discussion}
\label{expsec}

We evaluated our data profiling tool for accuracy and generalizability of rules for extracting UBSR concepts and higher-order semantic concepts from multi-lingual code data sets.

\textbf{Model Used:} To generate the rules for UBSR and higher-order concept profiler, we used the \texttt{Llama-3-70B-Instruct} model \cite{meta-llama3-instruct-70b}. Our UBSR rule generation relies on strong pattern matching capabilities for understanding the structure of known language ASTs and translating them to rules for unknown languages. In contrast, our semantic rule generation relies on pretraining knowledge related to libraries and their semantics. Given these requirements, our system is able to benefit from a frontier LLM such as \texttt{Llama-3-70B-Instruct}.

% Additionally, its superior instruction-following capabilities ensure that it adheres faithfully to prompt instructions, which is important for our offline rule generation. 
% \textcolor{red}{why eval tasks?}
% % For the higher-order concept profiler, we also evaluated the performance of the proposed system on other frontier models -- IBM Granite-34B-Code-Instruct, Meta CodeLlama-34B-Code-Instruct, and DeepSeek AI Deepseek-Coder-33B-Instruct.
% % \textcolor{green}{AE: Either do not mention Granite or do not mention discarded model alternatives anywhere to avoiding dissing on Granite ? This also helps preserve anonymity. --- in the last meeting we decided to run one line.}
% However, Llama-3-70B-Instruct demonstrated the best performance in all cases and hence, we selected this model for prompt evaluation.

\textbf{Datasets:} For UBSR concept extraction, we focused on programming languages from the aforementioned three syntactic paradigms, all of which support package, comment, and function concept types. The prompt exemplars for rule generation in the offline path comprised minimal code snippets from two programming languages (marked by * in Table~\ref{ubsr-support}) from each paradigm. For the evaluation of the online path, we used open datasets from GitHub \cite{cpprepo,scalarepo,typescriptrepo} in each test language, and matched the extracted concepts with the language-specific concept extractors for each language programmatically and manually.

For testing the efficacy of rule generation for semantic concept extraction \ref{eval:sem}, we generated an evaluation dataset comprising packages in different programming languages and catering to different functional domains. We used libraries.io \cite{librariesio}, an open-source repository of open-source packages available on the internet. Owing to the limited support of libraries.io and the noisy nature of its database, we limited ourselves to the top 15 languages (written in bold in Table \ref{ubsr-support}) among the 21 languages supported by the UBSR generator for evaluation. For each of these languages, we queried libraries.io using the language and semantic concepts listed in the aforementioned prompt and obtained a list of 4-8 packages per concept. For few-shot examples, we selected packages with their semantic concept mapping from Python and Java (one package per concept).

Finally, to demonstrate the output of the data profiler \ref{eval:e2e}, we created a synthetic dataset containing multi-lingual (languages written in bold in Table~\ref{ubsr-support}) code snippets with packages, comments, and functions and based on the aforementioned semantic concepts. 

% based on user-controlled concept classes spanning  UBSR concepts, sampled from real-world libraries across multiple programming languages from \ref{ubsr-support}. This ensures a diverse and balanced code data set spanning syntactic and semantic concepts across different programming languages. 

%textcolor{green}{AE: Take out, we don't need to explain the "how" of generating synth data to this level of detail.}Each snippet is assigned a weight based on an equal-weightage distribution for each language-class pair, which determines the number of snippets containing packages that represent semantic concepts. 

% \vspace{-10pt}
\subsection{Evaluation for Syntactic Concept Extraction}
\label{eval}
\textit{Robustness of LLM-Generated Rules Applied To Open Datasets:} In this experiment, we evaluated the robustness of the LLM-generated base syntactic rules in accurately extracting concepts from open datasets in diverse languages across each syntactic paradigm: C++ (C-like), Scala (Functional), and TypeScript (Scripting). The extracted UBSR concept counts from the raw code matched the manually generated AST-specific concept counts for package, comment, and function concepts, as shown in Table \ref{tab:validation_with_match_rate} as the UBSR count divided by the raw code count. This indicates that when the rules were applied, they extracted the concepts without any errors, achieving 100\% precision and recall. This establishes the criterion for evaluating the robustness of the prompt in the offline path, specifically for generating base syntactic rules across different languages and syntactic paradigms.

\begin{table}[h]
\scriptsize
\centering
\caption{Validation of Extracted Syntactic Concepts for Different Languages on Open Datasets \cite{cpprepo,scalarepo,typescriptrepo}. Rates are shown as extracted counts (UBSR / raw code).}
\begin{tabular}{|p{2cm}|p{2cm}|p{2.2cm}|p{2.2cm}|p{2.2cm}|}
\hline
\textbf{Language} & \textbf{Code Files} & \multicolumn{3}{c|}{\textbf{Extracted Concept Count in the Open Datasets (UBSR / raw code)}} \\
\cline{3-5}
 &  & \textbf{Package} & \textbf{Comment} & \textbf{Function} \\
\hline
C++              & 55 & 87/87 & 392/392 & 104/104 \\
Scala            & 44  & 26/26 & 30/30  & 37/37 \\
TypeScript       & 205 & 131/131 & 363/363 & 23/23 \\
\hline
\end{tabular}
\label{tab:validation_with_match_rate}
\end{table}

\textit{Impact of Syntactic Paradigm Consistency on Rule Generation Accuracy:} In this evaluation, we assessed the accuracy of LLM-generated syntactic rules, measured as the proportion of test languages for which correct rules were generated. We compared two scenarios: (a) prompt exemplars and test languages selected from the same paradigm, and (b) prompt exemplars and test languages selected from different paradigms. For rule extraction across all concepts, maintaining the same syntactic paradigm between prompt and test examples resulted in accuracy improvements ranging from 2.3\% to 125\% (or 2.25x) compared to the cross-paradigm scenario.

\begin{figure}[h]
        \centering
        \includegraphics[width=0.49\linewidth]{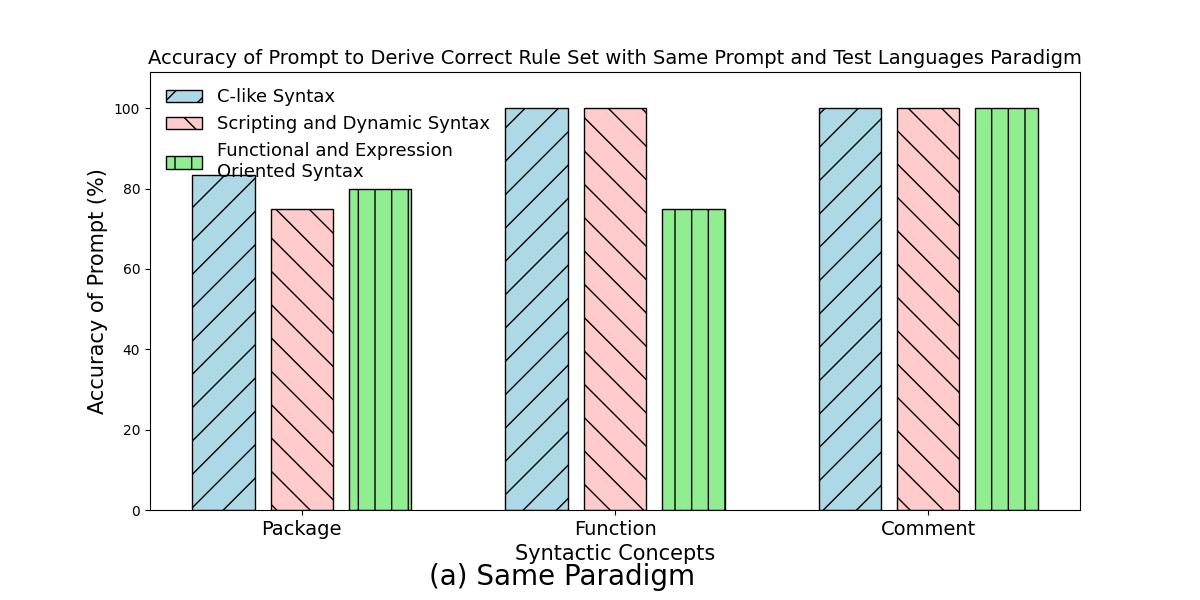} 
        \includegraphics[width=0.49\linewidth]{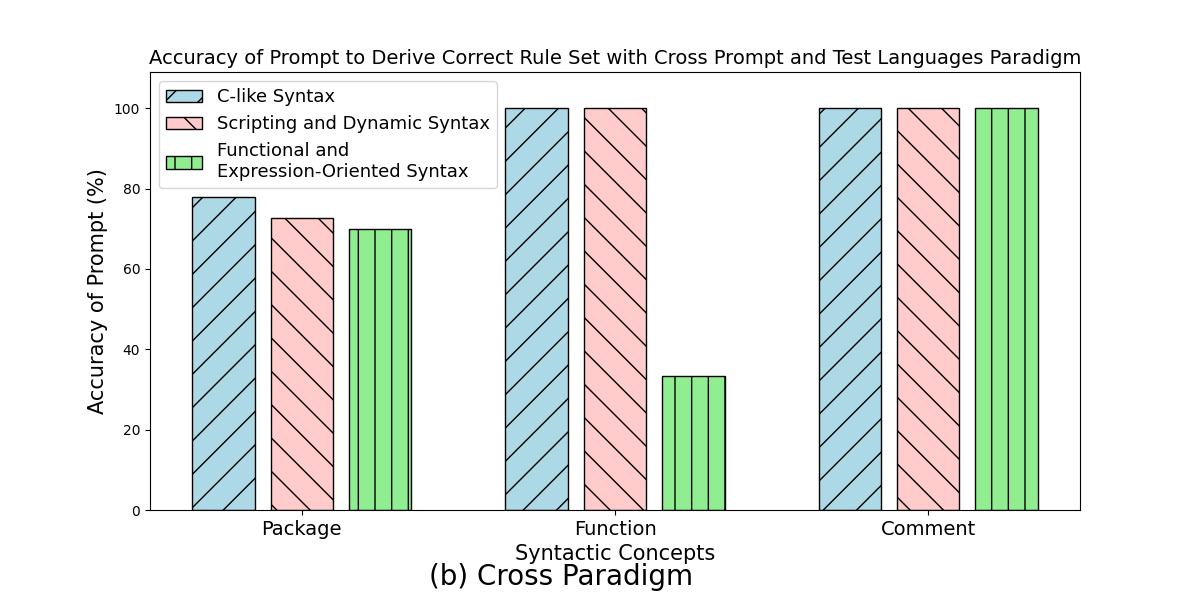}
        \caption{Rule Generation Accuracy with Prompt and Test Examples from (a) Same Paradigm (b) Cross Paradigm}
      \label{fig:mainfig}
\end{figure}

\iffalse
\begin{figure}[h]
\centering
    \begin{minipage}{.66\textwidth}
        \centering
        \hspace{-30pt}
        \includegraphics[width=0.53\linewidth]{figures/same-paradigm.png} 
        \hspace{-16pt}
        \includegraphics[width=0.53\linewidth]{figures/cross-paradigm.png}
        %\vspace{-14pt}
        \caption{Rule Generation Accuracy with Prompt and Test\\Examples from (a) Same Paradigm (b) Cross Paradigm}
      \label{fig:mainfig}
    \end{minipage}
    \hspace{-10pt}
    \begin{minipage}{.34\textwidth}
        \centering
        \hspace{-40pt}
        \includegraphics[width=\linewidth]{figures/ast-pruning.png}
        %\vspace{-14pt}
        \caption{Impact of AST Pruning on Token Reduction in Prompts}
        \label{fig:ast-pruning-graph-label}
    \end{minipage}
%\vspace{-12pt}
\end{figure}
\fi

%\textcolor{green}{Can we use the exact ubsr concept names e.g. ubsr-package etc in the explanations below ? }

For the package concept type, the rule extraction within the same paradigm showed an accuracy improvement of 7.14\% for c-like syntax, 3.12\% in scripting and dynamic syntax, and 14.28\% in functional and expression-oriented syntax paradigms over the cross-paradigm scenario. For the function concept type, the accuracy improvement was most pronounced in the functional and expression-oriented syntax paradigm, with a significant increase of 125\% (or 2.25x) when the prompt and test languages shared the same syntactic paradigm. For the concept comment, accuracy remained consistently high across all paradigms, likely due to the universal nature of comments, with no notable difference between same- and cross-paradigm scenarios. Overall, the same paradigm scenario exhibits a mean accuracy of 90.33\% for syntactic concept extraction rules across syntactic constructs and languages
These results show that maintaining syntactic paradigm consistency between prompt and test examples boosts the LLM’s accuracy in rule generation, particularly for complex concepts like packages and functions.

%\text{Improvement} = \frac{\text{Same paradigm accuracy} - \text{Cross paradigm accuracy}}{\text{Cross paradigm accuracy}} \times 100

\textit{Impact of AST Pruning on Token Reduction in LLM Prompts: }In this evaluation, we demonstrate that pruned ASTs are effective substitutes for unpruned trees, as pruning reduces AST size without compromising the LLM’s ability to generate correct syntactic rules. We illustrate this by using per-paradigm prompts for the package extraction task, with distinct prompt exemplars for each paradigm. This evaluation generalizes to other syntactic concepts since the pruning techniques are the same across UBSR concept types. From Figure \ref{fig:ast-pruning-graph-label}, we observe that concept-level pruning consistently results in the smallest AST size, reducing the token count on average by 64.03\% (2.78x) compared to no pruning, 50.06\% (2.00x) compared to depth-3 pruning, 27.25\% (1.37x) compared to depth-2, and 3.87\% (1.04x) compared to depth-1 pruning across all paradigms. These results highlight the efficiency of concept-level pruning in minimizing token overhead while maintaining rule generation accuracy.

\begin{figure}[h]
    %\vspace{-12pt}
    \centering
    \includegraphics[width=0.5\linewidth]{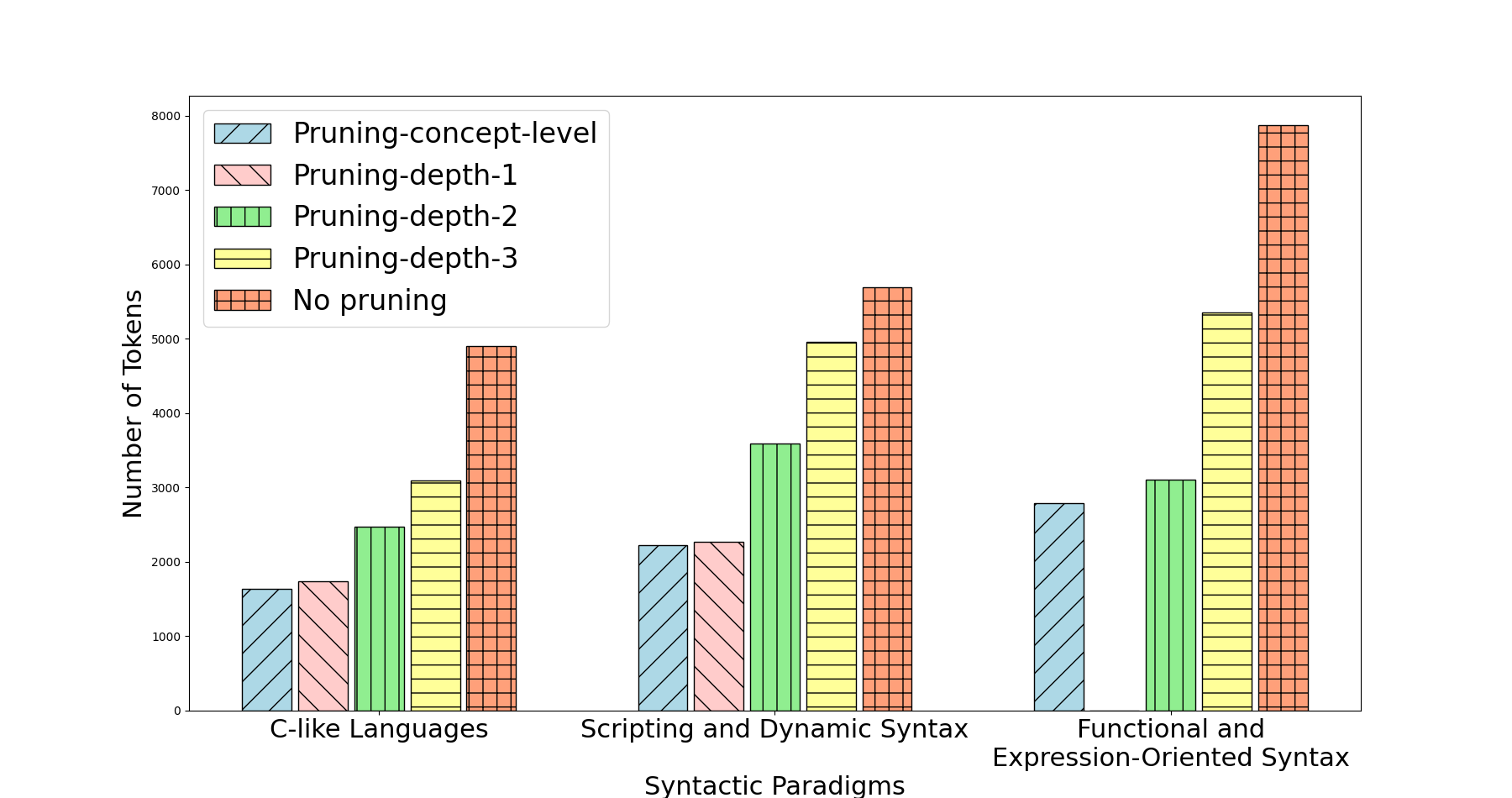}
    %\vspace{-12pt}
    \caption{Impact of AST Pruning on Token Reduction in LLM Prompts}
    \label{fig:ast-pruning-graph-label}
    %\vspace{-12pt}  
\end{figure}
% \vspace{-10pt}
\subsection{Evaluation of Higher Order Concept Profiler}\label{eval:sem}
% \textcolor{green}{AE : Do we want to mention that we don't explicitly measure accuracy of high-order syntactic constructs since, unlike semantic rules, they are deterministic rules derived from low-order concepts ? Also change the title of the section to "Evaluation of High-Order Profiling" ?  }
In the case of the higher-order concept profiler, the syntactic profiler utilizes deterministic rules to generate the higher-order syntactic concepts from the base syntactic concepts. Hence, we do not explicitly measure the accuracy of these constructs as it only depends on the accuracy of the rules and the base constructs. The semantic profiler, on the other hand, depends on rules generated using LLM and hence needs to be evaluated.

The core of the semantic profiler is the semantic rules database which is generated with the help of LLM prompting. Therefore, to evaluate the efficacy of the semantic profiler, we primarily focus on evaluating the accuracy of the semantic mapping prompt.

\begin{figure}[h]
\centering
\begin{minipage}{.5\textwidth}
  \centering
  \includegraphics[scale=0.1]{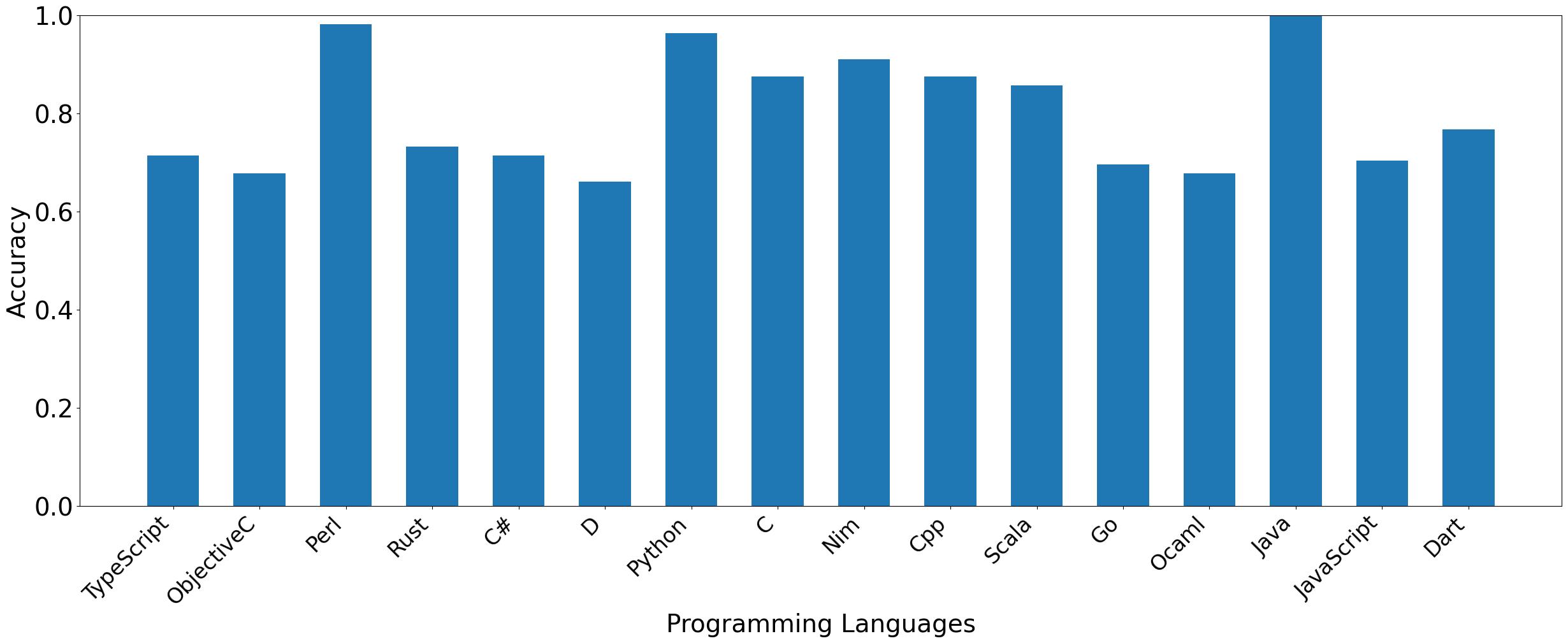}
   %\vspace{-10pt}
  \caption{Prompt Accuracy Across Various Languages.}
  \label{fig:lang}
\end{minipage}%
\begin{minipage}{.5\textwidth}
  \centering
  \includegraphics[scale=0.1]{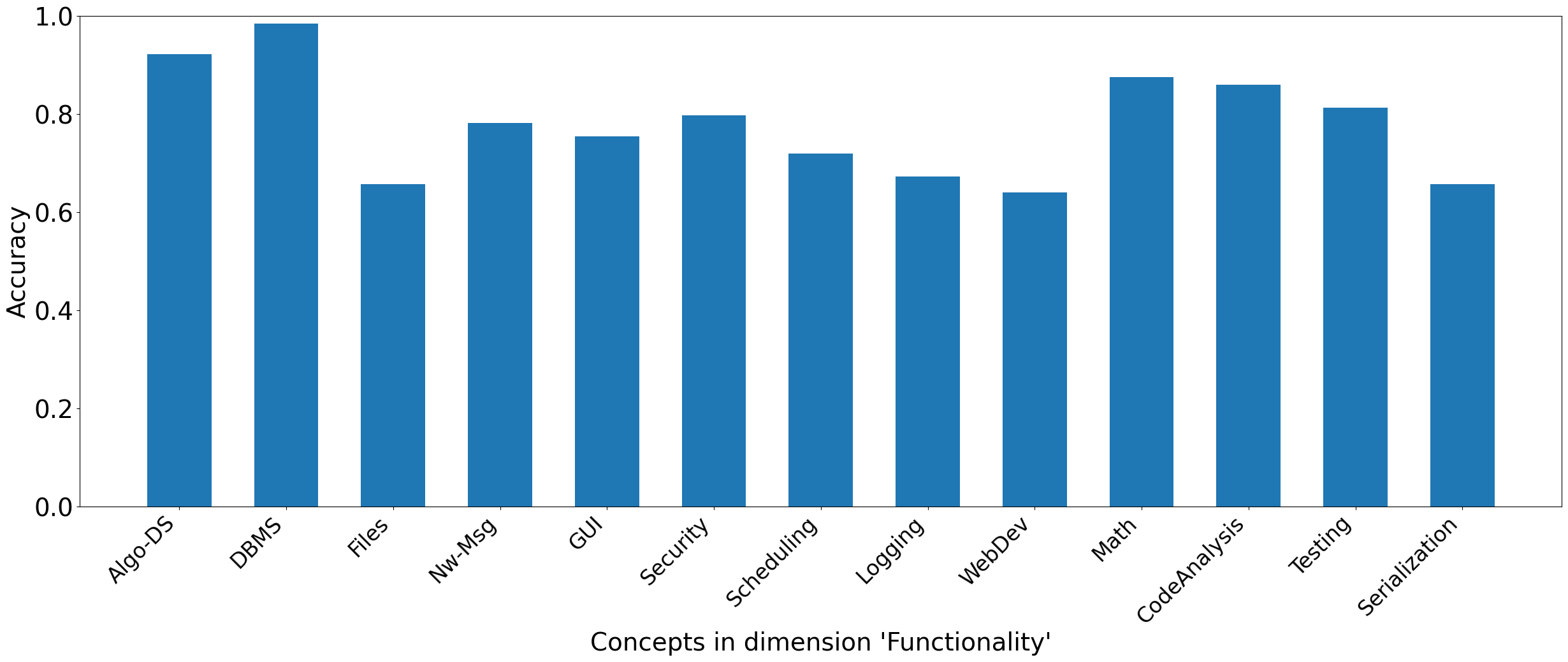}
  %\vspace{-11pt}
  \caption{Prompt Accuracy Across Various Concepts.}
  \label{fig:conc}
\end{minipage}
%\vspace{-5pt}
\end{figure}

We evaluated the efficacy of this prompt by testing it on the Llama-3 model. We used APIs exposed by the model to programmatically submit the prompt and read the response, which is then processed to obtain the output in the desired tabular format. Due to the limitation in the maximum number of tokens that are generated in each response, we input the list of packages in batches of 30 for categorization by the LLM. On obtaining the output, we compare it with the ground truth generated using libraries.io and calculate the accuracy, defined as the ratio of the number of packages that are classified by the model to be of the same semantic concept as that in the ground truth to the total number of packages. We measured the accuracy of prompt results and analysed their variation across various programming languages as well as the chosen semantic concepts. The plots are presented in Figures \ref{fig:lang} and \ref{fig:conc} and discussed below.

\textit{Accuracy of prompt results across languages:} From Figure \ref{fig:lang}, we observed that the proposed prompt demonstrates an accuracy of atleast 65-100\% across the various programming languages. For both the prompt languages Java and Python, the accuracy is almost 98 -- 100 \%, whereas among the test languages, the best performance is observed in case of Perl, followed closely by Nim, Cpp, C and Scala. 
% The language specific performance depends on multiple factors such as whether and how well the base model has been trained on those packages or languages.
%\textcolor{green}{AE : I think we should take the previous line out unless we know for sure that training data the base models used correlates with the model accuracy we observed}
Digging further into the results, we observed that packages with more intuitive names are better categorised by the model.

\textit{Accuracy of prompt results across concepts:} From Figure \ref{fig:conc}, we observed that the proposed prompt demonstrates an accuracy of at least 62-98\% across the semantic concepts in the functionality dimension. The top 5 best-performing concepts in this case are Database, Algorithms, Mathematics, Code Analysis, and Testing. An interesting observation in this case was that the chosen names of concepts play a significant role in determining the accuracy of the prompt results. Specifically, the concepts need to be chosen such that these are as semantically non-overlapping to each other as possible ensuring there are no conceptual overlaps.

Another issue that we observed in the evaluation of the prompt results is that individual programming language packages often serve multiple purposes\footnote{e.g., jwt-cpp is a Cpp package used for creating and validating JSON Web Tokens. The libraries.io ground truth data classifies this as a Web Development package, whereas the LLM classifies it as a security package. NetworkEye is another such package in Objective-C used for debugging and monitoring HTTP requests, thereby supporting both functionalities of logging and monitoring as classified in ground truth and networking and messaging as classified by the model.}. Such ambiguities also played a significant role in determining the performance of the LLM on the prompt. This issue can be resolved to a certain extent by classifying the packages into multiple relevant concepts, instead of just one, which is a possible future research direction.

\subsection{End-to-End Code Data Profiling Report}\label{eval:e2e}
\begin{figure}[h]
    \centering
    \includegraphics[width=0.8\linewidth]{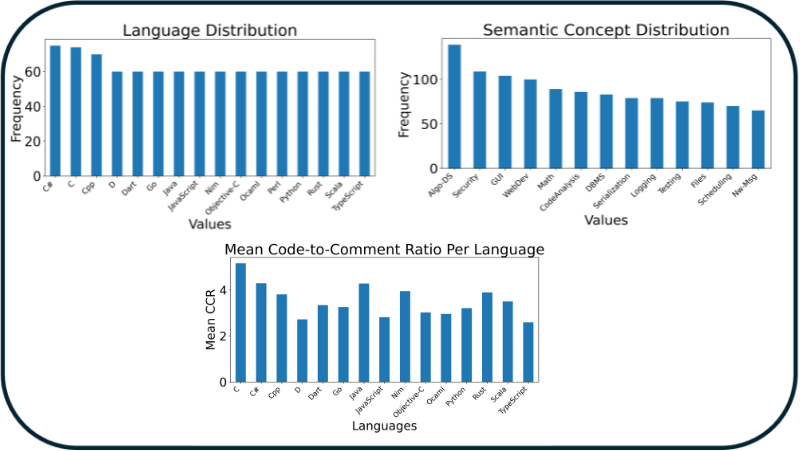}
    % \begin{minipage}{.45\textwidth}
    % \centering
    %     \includegraphics[width=1.0\linewidth]{figures/concept_counts_model.png}
    % \end{minipage}
    % \begin{minipage}{.45\textwidth}
    %  \centering
    %     \includegraphics[width=1.0\linewidth]{figures/lang_counts_model.png}
    % \end{minipage}
    % \begin{minipage}{.4\textwidth}
    %  \centering
    %     \includegraphics[width=\linewidth]{figures/ccr_counts_model.png}
    % \end{minipage}
    % \vspace{-10pt}
    \caption{End-to-End Code Data Profiler Output.}
    \label{fig:e2e}
    %\vspace{-10pt}
\end{figure}

In this section, we present the higher-order syntactic and semantic statistical  derived from profiling the synthetic dataset. A sample output of the data profiler is shown in Fig. \ref{fig:e2e}. Specifically, the output contains the distribution of programming languages, the higher-order syntactic concept of code-to-comment ratio, and semantic concepts under `functionality' across the synthetic dataset.

% we derived the higher-order syntactic concept called code-to-comment ratio from the base syntactic concepts present in the UBSR. This metric can be used as a criterion for data filtering. On the other hand, for semantic profiling, we focused on the distribution of libraries and semantic classes present within the dataset. The Library Distribution plot shown in the figure presents the top 50 libraries in the dataset and the number of files with these libraries, thereby offering a clear picture of the dominant technologies and tools within the dataset. The Class Distribution plot presented in the figure visualizes the most frequently used classes and the number of files containing each class. By evaluating the richness and diversity of concepts, users can effectively guide dataset curation for LLM customization, ensuring that the semantic distribution aligns with the downstream use case.

% \begin{figure}
%     \centering
%     \includegraphics[width=0.7\linewidth]{figures/e2e.png}
%     \caption{End-to-End Code Data Profiler Output}
%     \label{fig:e2e}
% \end{figure}

%\begin{itemize}
%    \item Synthetic Data set generation
%    \item About the script, weights
%    \item Experimental flow
%    \item Report
%    \item Semantic: Code to comment, etc
%    \item Semantic:  lang and concept dist.
%\end{itemize}

\section{Conclusion and Future Work}
\label{conclusion}
In this work, we motivate, build, and evaluate an extensible data profiling tool for characterizing the properties of multi-lingual code data sets in terms of user-defined syntactic and semantic concepts. At the foundation of our approach is the capability to convert unstructured multi-lingual code data into a language-agnostic UBSR and layer customizable higher-order syntactic and semantic concepts on top of the UBSR. 
% Through a hybrid approach that combines LLM-based rule generation in the offline phase with deterministic rule application in the online phase, our system can achieve high accuracy \textcolor{red}{(X percent and Y percent)} in real-world multi-lingual multi-paradigmatic datasets, validating the robustness and generalizability of the approach.
Through a hybrid approach that combines LLM-based rule generation in the offline phase with deterministic rule application in the online phase, our system generates 100\% accurate syntactic rules in real-world multi-lingual, multi-paradigmatic datasets. Our offline UBSR extraction approach exhibits a mean accuracy of 90.33\% for syntactic concept extraction rules across syntactic constructs and languages and reduces token overhead by a factor of 2.78x, whereas our semantic profiler demonstrates a mean accuracy of 77/80\% across languages/concepts.

The unified tabular representation for code samples can be used as the basis for a variety of downstream data-processing tasks. 
% In data valuation, the value of each data sample and the over-all data set can be assessed along metrics related to data quality, relevance relative to target objectives etc. 
The syntactic and semantic properties extracted via the profiler are a natural foundation for deriving use-case customized metrics for data valuation. The tabular representations extracted can also be used as the basis for shared multi-lingual curation operators for filtering or ranking code data points. These aspects will be explored in future. 

% \textbf{Data Availability}: 

%\section{Data Availability}
% To promote open science, we have made our code and data publicly available at the following link: \url{https://doi.org/10.5281/zenodo.13757379}.

% \begin{acks}
% \end{acks}  

%\bibliographystyle{ACM-Reference-Format}
\bibliographystyle{plain}  % or another style like acmart, IEEEtran
\bibliography{sample-base}

\begin{thebibliography}{10}

\bibitem{abedjan2015profiling}
Ziawasch Abedjan, Lukasz Golab, and Felix Naumann.
\newblock Profiling relational data: a survey.
\newblock {\em The VLDB Journal}, 24:557--581, 2015.

\bibitem{snowball2000}
Eugene Agichtein and Luis Gravano.
\newblock Snowball: extracting relations from large plain-text collections.
\newblock In {\em Proceedings of the Fifth ACM Conference on Digital
  Libraries}, New York, NY, USA, 2000. Association for Computing Machinery.

\bibitem{ahmad2021unifiedpretrainingprogramunderstanding}
Wasi~Uddin Ahmad, Saikat Chakraborty, Baishakhi Ray, and Kai-Wei Chang.
\newblock Unified pre-training for program understanding and generation, 2021.

\bibitem{meta-llama3-instruct-70b}
Meta AI.
\newblock Meta llama 3: 8b instruct model.
\newblock \url{https://huggingface.co/meta-llama/Meta-Llama-3-70B-Instruct},
  2024.
\newblock Accessed: 2024-09-01.

\bibitem{scalarepo}
The Algorithms.
\newblock Scala algorithms.
\newblock \url{https://github.com/TheAlgorithms/Scala.git}.
\newblock Accessed on September 2024.

\bibitem{typescriptrepo}
The Algorithms.
\newblock Typescript algorithms.
\newblock \url{https://github.com/TheAlgorithms/TypeScript.git}.
\newblock Accessed on September 2024.

\bibitem{arora2023languagemodelsenablesimple}
Simran Arora, Brandon Yang, Sabri Eyuboglu, Avanika Narayan, Andrew Hojel,
  Immanuel Trummer, and Christopher Ré.
\newblock Language models enable simple systems for generating structured views
  of heterogeneous data lakes, 2023.

\bibitem{47967}
Eric Breck, Marty Zinkevich, Neoklis Polyzotis, Steven Whang, and Sudip Roy.
\newblock Data validation for machine learning.
\newblock In {\em Proceedings of SysML}, 2019.

\bibitem{treesitter}
Max Brunsfeld.
\newblock Tree-sitter: An incremental parsing system for programming tools,
  2023.
\newblock Accessed: 2024-08-24.

\bibitem{shredder12}
Kuang Chen, Akshay Kannan, Yoriyasu Yano, Joseph Hellerstein, and Tapan Parikh.
\newblock Shreddr: Pipelined paper digitization for low-resource organizations.
\newblock {\em Proceedings of the 2nd ACM Symposium on Computing for
  Development, DEV 2012}, 05 2012.

\bibitem{chen2024seeddomainspecificdatacuration}
Zui Chen, Lei Cao, Sam Madden, Tim Kraska, Zeyuan Shang, Ju~Fan, Nan Tang,
  Zihui Gu, Chunwei Liu, and Michael Cafarella.
\newblock Seed: Domain-specific data curation with large language models, 2024.

\bibitem{holisticcleaning}
Xu~Chu, Ihab~F. Ilyas, and Paolo Papotti.
\newblock Holistic data cleaning: Putting violations into context.
\newblock In {\em 2013 IEEE 29th International Conference on Data Engineering
  (ICDE)}, pages 458--469, 2013.

\bibitem{côté2024datacleaningmachinelearning}
Pierre-Olivier Côté, Amin Nikanjam, Nafisa Ahmed, Dmytro Humeniuk, and Foutse
  Khomh.
\newblock Data cleaning and machine learning: A systematic literature review,
  2024.

\bibitem{LLMStructCuration3}
Xiang Deng, Huan Sun, Alyssa Lees, You Wu, and Cong Yu.
\newblock Turl: Table understanding through representation learning, 2020.

\bibitem{du2024codegragextractingcomposedsyntax}
Kounianhua Du, Renting Rui, Huacan Chai, Lingyue Fu, Wei Xia, Yasheng Wang,
  Ruiming Tang, Yong Yu, and Weinan Zhang.
\newblock Codegrag: Extracting composed syntax graphs for retrieval augmented
  cross-lingual code generation, 2024.

\bibitem{epperson2023deadalivecontinuousdata}
Will Epperson, Vaishnavi Gorantla, Dominik Moritz, and Adam Perer.
\newblock Dead or alive: Continuous data profiling for interactive data
  science, 2023.

\bibitem{feng2020codebertpretrainedmodelprogramming}
Zhangyin Feng, Daya Guo, Duyu Tang, Nan Duan, Xiaocheng Feng, Ming Gong, Linjun
  Shou, Bing Qin, Ting Liu, Daxin Jiang, and Ming Zhou.
\newblock Codebert: A pre-trained model for programming and natural languages,
  2020.

\bibitem{octoverse:top_programming_languages}
GitHub.
\newblock Top programming languages of 2022, 2022.
\newblock Accessed: 2024-09-12.

\bibitem{gong2024astt5structureawarepretrainingcode}
Linyuan Gong, Mostafa Elhoushi, and Alvin Cheung.
\newblock Ast-t5: Structure-aware pretraining for code generation and
  understanding, 2024.

\bibitem{guo2021graphcodebertpretrainingcoderepresentations}
Daya Guo, Shuo Ren, Shuai Lu, Zhangyin Feng, Duyu Tang, Shujie Liu, Long Zhou,
  Nan Duan, Alexey Svyatkovskiy, Shengyu Fu, Michele Tufano, Shao~Kun Deng,
  Colin Clement, Dawn Drain, Neel Sundaresan, Jian Yin, Daxin Jiang, and Ming
  Zhou.
\newblock Graphcodebert: Pre-training code representations with data flow,
  2021.

\bibitem{huang2024cocoonsemantictableprofiling}
Zezhou Huang and Eugene Wu.
\newblock Cocoon: Semantic table profiling using large language models, 2024.

\bibitem{metareader}
Hassan Jannah.
\newblock Metareader: A dataset meta-exploration and documentation tool, 12
  2014.

\bibitem{kim2021codepredictionfeedingtrees}
Seohyun Kim, Jinman Zhao, Yuchi Tian, and Satish Chandra.
\newblock Code prediction by feeding trees to transformers, 2021.

\bibitem{lachaux2020unsupervisedtranslationprogramminglanguages}
Marie-Anne Lachaux, Baptiste Roziere, Lowik Chanussot, and Guillaume Lample.
\newblock Unsupervised translation of programming languages, 2020.

\bibitem{li2023starcodersourceyou}
Raymond Li, Loubna~Ben Allal, and Yangtian~Zi et~al.
\newblock Starcoder: may the source be with you!, 2023.

\bibitem{lin2024accurateefficientdocumentanalytics}
Yiming Lin, Madelon Hulsebos, Ruiying Ma, Shreya Shankar, Sepanta Zeigham,
  Aditya~G. Parameswaran, and Eugene Wu.
\newblock Towards accurate and efficient document analytics with large language
  models, 2024.

\bibitem{Liu_2023}
Mingxuan Liu, Siqi Li, Han Yuan, Marcus Eng~Hock Ong, Yilin Ning, Feng Xie,
  Seyed~Ehsan Saffari, Yuqing Shang, Victor Volovici, Bibhas Chakraborty, and
  Nan Liu.
\newblock Handling missing values in healthcare data: A systematic review of
  deep learning-based imputation techniques.
\newblock {\em Artificial Intelligence in Medicine}, 142:102587, August 2023.

\bibitem{raha2019}
Mohammad Mahdavi, Ziawasch Abedjan, Raul Fernandez, Samuel Madden, Mourad
  Ouzzani, Michael Stonebraker, and Nan Tang.
\newblock Raha: A configuration-free error detection system.
\newblock 01 2019.

\bibitem{LLMStructCuration1}
Yinan Mei, Shaoxu Song, Chenguang Fang, Haifeng Yang, Jingyun Fang, and Jiang
  Long.
\newblock Capturing semantics for imputation with pre-trained language models.
\newblock In {\em 2021 IEEE 37th International Conference on Data Engineering
  (ICDE)}, pages 61--72, 2021.

\bibitem{mishra2024granite}
Mayank Mishra, Matt Stallone, Gaoyuan Zhang, Yikang Shen, Aditya Prasad,
  Adriana~Meza Soria, Michele Merler, Parameswaran Selvam, Saptha Surendran,
  Shivdeep Singh, et~al.
\newblock Granite code models: A family of open foundation models for code
  intelligence.
\newblock {\em arXiv preprint arXiv:2405.04324}, 2024.

\bibitem{LLMStructCuration2}
Avanika Narayan, Ines Chami, Laurel Orr, Simran Arora, and Christopher Ré.
\newblock Can foundation models wrangle your data?, 2022.

\bibitem{librariesio}
Andrew Nesbitt.
\newblock Libraries.io: The open source discovery service, Accessed:
  2024-09-12.

\bibitem{papadakis2020surveyblockingfilteringtechniques}
George Papadakis, Dimitrios Skoutas, Emmanouil Thanos, and Themis Palpanas.
\newblock A survey of blocking and filtering techniques for entity resolution,
  2020.

\bibitem{antlr}
Terence Parr.
\newblock Antlr - another tool for language recognition, Accessed: 2024-09-12.

\bibitem{poesia2022synchromeshreliablecodegeneration}
Gabriel Poesia, Oleksandr Polozov, Vu~Le, Ashish Tiwari, Gustavo Soares,
  Christopher Meek, and Sumit Gulwani.
\newblock Synchromesh: Reliable code generation from pre-trained language
  models, 2022.

\bibitem{babel}
Babelfish Project.
\newblock Babelfish uast, Accessed: 2024-09-12.

\bibitem{kythe}
Kythe Project.
\newblock Kythe - a pluggable, (mostly) language-agnostic ecosystem for
  building tools that work with code., Accessed: 2024-09-12.

\bibitem{rekatsinas2017holocleanholisticdatarepairs}
Theodoros Rekatsinas, Xu~Chu, Ihab~F. Ilyas, and Christopher Ré.
\newblock Holoclean: Holistic data repairs with probabilistic inference, 2017.

\bibitem{roziere2021dobfdeobfuscationpretrainingobjective}
Baptiste Roziere, Marie-Anne Lachaux, Marc Szafraniec, and Guillaume Lample.
\newblock Dobf: A deobfuscation pre-training objective for programming
  languages, 2021.

\bibitem{rozière2024codellamaopenfoundation}
Baptiste Rozière, Jonas Gehring, and Fabian~Gloeckle et~al.
\newblock Code llama: Open foundation models for code, 2024.

\bibitem{scott2016programming}
Michael~L. Scott.
\newblock {\em Programming Language Pragmatics}.
\newblock Morgan Kaufmann, 4th edition, 2016.

\bibitem{10.1145/3329486.3329499}
Vraj Shah and Arun Kumar.
\newblock The ml data prep zoo: Towards semi-automatic data preparation for ml.
\newblock DEEM'19, New York, NY, USA, 2019. Association for Computing
  Machinery.

\bibitem{streamlit}
{Streamlit Inc.}
\newblock Streamlit: A faster way to build and share data apps.
\newblock \url{https://streamlit.io/}, 2024.
\newblock Accessed: 2024-08-29.

\bibitem{articletreegen}
Zeyu Sun, Qihao Zhu, Yingfei Xiong, Yican Sun, Lili Mou, and Lu~Zhang.
\newblock Treegen: A tree-based transformer architecture for code generation.
\newblock {\em Proceedings of the AAAI Conference on Artificial Intelligence},
  34:8984--8991, 04 2020.

\bibitem{tipirneni2024structcoderstructureawaretransformercode}
Sindhu Tipirneni, Ming Zhu, and Chandan~K. Reddy.
\newblock Structcoder: Structure-aware transformer for code generation, 2024.

\bibitem{tsai2024codelessalignmore}
Yun-Da Tsai, Mingjie Liu, and Haoxing Ren.
\newblock Code less, align more: Efficient llm fine-tuning for code generation
  with data pruning, 2024.

\bibitem{cpprepo}
Sinair V.
\newblock Cpp tutorial samples.
\newblock \url{https://github.com/sinairv/Cpp-Tutorial-Samples}.
\newblock Accessed on September 2024.

\bibitem{wang-etal-2021-codet5}
Yue Wang, Weishi Wang, Shafiq Joty, and Steven~C.H. Hoi.
\newblock {C}ode{T}5: Identifier-aware unified pre-trained encoder-decoder
  models for code understanding and generation.
\newblock In {\em Proceedings of the 2021 Conference on Empirical Methods in
  Natural Language Processing}, pages 8696--8708, Online and Punta Cana,
  Dominican Republic, November 2021. Association for Computational Linguistics.

\bibitem{genai2024}
IBM Watson.
\newblock Genai: A generative ai python library.
\newblock \url{https://pypi.org/project/genai/}, 2024.
\newblock Version 0.0.221.

\bibitem{wei2023chainofthoughtpromptingelicitsreasoning}
Jason Wei, Xuezhi Wang, Dale Schuurmans, Maarten Bosma, Brian Ichter, Fei Xia,
  Ed~Chi, Quoc Le, and Denny Zhou.
\newblock Chain-of-thought prompting elicits reasoning in large language
  models, 2023.

\bibitem{wu2024learningextractstructuredentities}
Haolun Wu, Ye~Yuan, Liana Mikaelyan, Alexander Meulemans, Xue Liu, James
  Hensman, and Bhaskar Mitra.
\newblock Learning to extract structured entities using language models, 2024.

\bibitem{wu2024structureawarefinetuningcodepretrained}
Jiayi Wu, Renyu Zhu, Nuo Chen, Qiushi Sun, Xiang Li, and Ming Gao.
\newblock Structure-aware fine-tuning for code pre-trained models, 2024.

\bibitem{yin2017syntacticneuralmodelgeneralpurpose}
Pengcheng Yin and Graham Neubig.
\newblock A syntactic neural model for general-purpose code generation, 2017.

\bibitem{zügner2021languageagnosticrepresentationlearningsource}
Daniel Zügner, Tobias Kirschstein, Michele Catasta, Jure Leskovec, and Stephan
  Günnemann.
\newblock Language-agnostic representation learning of source code from
  structure and context, 2021.

\end{thebibliography}

\appendix

\end{document}